\documentclass[10pt, twocolumn,english,prd,showpacs,nofootinbib]{revtex4}
\usepackage[latin9]{inputenc}
\usepackage{graphicx}
\usepackage{amsmath}
\usepackage{amssymb}
\usepackage{amsfonts}
\usepackage{lscape}
\usepackage{epsfig}

\usepackage{color}
\usepackage{bm}

\providecommand{\tabularnewline}{\\}

\usepackage{babel}
\def\be{\begin{equation}}
\def\ee{\end{equation}}
\def\bea{\begin{eqnarray}}
\def\eea{\end{eqnarray}}
\def\cs{c_{\rm s}^2}
\def\ce{c_{\rm eff}^2}
\def\cv{c_{\rm v}^2}
\def\ca{c_{\rm a}^2}
\def\Om{\Omega_{m,0}}
\def\ceff{c_{\rm eff}^2}
\def\lzr{\left(z\right)}

\newcommand{\M}{\mathcal{M}}

\begin{document}

\title{Can dark energy viscosity be detected with the Euclid survey?}

\date{\today}

\author{Domenico Sapone}

\email{domenico.sapone@uam.es}

\affiliation{Departamento de F\'isica Te\'orica and Instituto de F\'isica Te\'orica, \\
 Universidad Aut\'onoma de Madrid IFT-UAM/CSIC,\\
 $28049$ Cantoblanco, Madrid, Spain}

\author{Elisabetta Majerotto}

\email{elisabetta.majerotto@uam.es}

\affiliation{Departamento de F\'isica Te\'orica and Instituto de F\'isica Te\'orica, \\
 Universidad Aut\'onoma de Madrid IFT-UAM/CSIC,\\
 $28049$ Cantoblanco, Madrid, Spain}

\author{Martin Kunz}

\email{martin.kunz@unige.ch}

\affiliation{D\'epartement de Physique Th\'eorique, Universit\'e de Gen\`eve, 24 quai
Ernest Ansermet, CH--1211 Gen\`eve 4, Switzerland}
\affiliation{African Institute for Mathematical Sciences, 6-8 Melrose Road, Muizenberg, Cape Town, South Africa}

\author{Bianca Garilli}

\email{bianca@lambrate.inaf.it}

\affiliation{IASF-Milano, INAF, Via Bassini, 15,\\ I--20133, Milano, Italy\\}

\begin{abstract}

Recent work has demonstrated that it is important to constrain the dynamics of cosmological perturbations,
in addition to the evolution of the background, if we want to distinguish among different 
models of the dark sector. Especially the anisotropic stress of the (possibly effective) dark energy
fluid has been shown to be an important discriminator between modified gravity and dark energy
models. In this paper we use approximate analytical solutions of the perturbation equations in
the presence of viscosity to study how the anisotropic stress affects the weak 
 lensing  and galaxy power spectrum. We then forecast how
  sensitive the photometric and spectroscopic Euclid surveys will
   be to both the speed of sound and the 
   viscosity of our effective dark energy fluid when using weak lensing tomography and the galaxy power spectrum. We find that Euclid alone can only constrain models with very small speed of sound and viscosity, while it will need the help of other observables in order to give interesting constraints on models with a sound speed close to one. This conclusion is also supported by the expected Bayes factor between models.

\end{abstract}


\maketitle

\def\be{\begin{equation}}
\def\ee{\end{equation}}
\def\bea{\begin{eqnarray}}
\def\eea{\end{eqnarray}}
\newcommand{\dep}{\delta p}

\global\long\global\long\global\long\def\cs{c_{s}^{2}}
 \global\long\global\long\global\long\def\rmd{{\rm d}}
 \global\long\global\long\global\long\def\ca{c_{a}^{2}}
 \global\long\global\long\global\long\def\de{\mathrm{DE}}
 \global\long\global\long\global\long\def\lkr{{\left(k\right)}}
 
\section{Introduction}

Since its discovery in 1998 by \cite{sn1,sn2}, the cause of cosmic acceleration has not been understood, 
despite all the observational and theoretical efforts in this direction (see e.g. \cite{sapone, kunzrev,lucabook} and references therein). 
Ironically, the best explanation from the observational point of view, i.e. a cosmological constant, 
is very little satisfactory from a theoretical perspective \cite{Carroll:2000fy}. Among the many alternative possible answers, 
one proposal is to modify the Einstein equations, either in 4 dimensions, like in 
scalar-tensor theories, or in 5D, as e.g. in the Dvali Gabadadze Porrati model \cite{dgp} or even 
in 6D \cite{dgp6, rugbyball}. All these theories, if rewritten as ``effective dark energy'' models (simply by 
moving the additional terms modifying Einstein equations from the geometry side to the matter side of the equations) exhibit a difference 
with respect to ordinary scalar field dark energy (DE): they possess anisotropic stress.
  
For this reason, and also because we still do not know what 
dark energy is really made of, in \cite{fingerprinting3} some 
of us studied an ``effective dark energy'' with anisotropic stress. 
The latter was modeled as in \cite{Hu:1998kj} with the help of a viscosity parameter, 
in addition to the speed of sound and equation of state parameters. This parameterization was used first in \cite{Koivisto:2005mm} in relation with dark energy. A further work by \cite{Mota:2007sz} analyzed the model with data from the cosmic microwave background radiation (CMB), large scale structure and type Ia supernovae, and showed that it is hard to constrain both $\cs$ and $\cv$ and that future data would not improve very much their measurement. 
In \cite{fingerprinting3} analytical equations describing the cosmological perturbations for this imperfect fluid dark energy were derived, following the lines of a previous paper \cite{fingerprinting1}, 
where the same was done for a model with no viscosity. 
Other recent work on anisotropic stress can be found in \cite{papers-anisotropy}.

In this paper we will use these analytical expressions to understand how well the galaxy clustering (GC)  and weak lensing (WL) measurements of the Euclid survey\footnote{http://www.euclid-ec.org/}  \cite{Cimatti:2009is, RedBook} will be able to constrain the viscosity of the dark energy fluid, together with its speed of sound.

The Euclid survey is a recently selected mission
of the ESA Cosmic Vision program, whose launch is planned
 for 2020. The reason why it is particularly apt to constrain
  imperfect fluid DE (or alternatively modified gravity models)
is that it will perform both a photometric survey to measure
 WL and a spectroscopic survey to measure the galaxy power 
 spectrum (and higher order functions). Both WL and the
  galaxy power spectrum are able to constrain not only the expansion history of the Universe, which depends on the matter density and the equation of state of DE, but also further observables like the growth of structure. From these additional observables it is possible in principle to measure the speed of sound and the viscosity parameter, and to constrain the evolution of the perturbations. These are precisely the distinctive features of such models. 

In this work we will not only forecast the total errors on the model parameters, by using the Fisher matrix formalism, but also analyze separately the different contributions, taking advantage of our analytical formulae. This will allow us to understand better which are the aspects of the WL and of the galaxy power spectrum that have the strongest impact.

We will find that the Euclid survey is marginally able to constrain the viscosity together with the speed of sound, as errors are of the order of $100\%$. This is due to the complexity of the model and to the smallness of the effects that we wish to detect.

By evaluating the forecasted Bayes' factor we will find moreover that there is (strong) evidence in favor of viscous dark energy, as compared to a dark energy model with the same sound speed but no viscosity, only when the fiducial viscosity and sound speed are very small (but not too small) and when both weak lensing and galaxy clustering are used. In the latter case, decisive evidence in favor of viscous dark energy can be reached if we reduce the maximum viscosity allowed by our flat prior to be less than $\simeq 10^{-1}$, otherwise the Occam's razor effect of the Bayes factor dominates and disfavours the presence of non-zero viscosity.

The plan of the paper is the following. In Sec. \ref{sec:main-equations} we briefly describe the model and give the basic equations together with the main formulae found in \cite{fingerprinting3}. We then analyze the different observables and evaluate analytically their sensitivity to the speed of sound and the viscosity parameter, in Sec. \ref{sec:derivatives}. Sec. \ref{sec:Fisher-forecasts} is devoted to the Fisher matrix forecasts on the errors for our model from the Euclid WL and GC surveys. We analyze our results taking into account the results of the previous section. We forecast the Bayesian evidence using our computed Fisher matrices in Sec. \ref{sec:model-comparison} and we finally conclude in Sec. \ref{sec:conclusions}.

\section{Approximate solutions and general behaviors} \label{sec:main-equations}

We start by describing the model we consider: an imperfect fluid dark energy with anisotropic stress.
In this section we give the basic equations, defining our notation and presenting the approximate 
analytical solution to the dark energy perturbation evolution found in \cite{fingerprinting3} 
(for more details please see the aforementioned paper).

\subsection{Definitions}

We consider scalar linear perturbations about a spatially flat Friedmann-Lemaitre-Robertson-Walker Universe, 
whose line element is, in conformal Newtonian gauge,
\be
ds^2 = a^2 \left[ -(1+2\psi) d\tau^2 + (1-2\phi) dx_i dx^i \right],
\ee
where $a$ is the scale factor, $\tau$ is the conformal time, $x_i$ are the spatial coordinates and 
$\psi$ and $\phi$ are the metric perturbations.
We take an imperfect fluid dark energy, with constant equation of state $w$, speed of sound $c_s$ 
and with an anisotropic stress component $\sigma$. The first order perturbation equations for this fluid are
\bea
\delta' &=& 3(1+w) \phi' - \frac{V}{Ha^2} - 3 \frac{1}{a}\left(\frac{\dep}{\rho}-w \delta \right) \label{delta} , \\
V' &=& -(1-3w) \frac{V}{a}+ \frac{k^2}{H a^2} \frac{\dep}{\rho}+(1+w) \frac{k^2}{Ha^2} \psi +\\ \nonumber
&-&(1+w)\frac{k^2}{Ha^2}\sigma , \label{v}
\eea
where $\delta$ and $V$ are the density contrast and the velocity perturbation, $\dep$ is the 
pressure perturbation, $H$ is the Hubble function, $\rho$ is the dark energy density and the prime refers to derivatives with respect to the scale factor $a$.
For the evolution of $\sigma$, we consider the model proposed by \cite{Hu:1998kj}:
\be
\sigma' + \frac{3}{a} \sigma = \frac{8}{3} \frac{\cv}{(1+w)^2} \frac{V}{a^2H} , \label{eq:sig}
\ee
 which, for $\cv = 1/3$, recovers the evolution of anisotropic stress for radiation up to the quadrupole
  and reduces 
to the case of a classical uncoupled scalar field, which has always $\sigma = 0$, when the 
viscosity parameter $\cv$ is set to zero\footnote{In the case where $\cv = 0$ the viscosity decays as $\sigma \sim a^{-3}$ 
even if it is initially non-zero and vanishes indeed rapidly.}.
Pressure perturbations are parameterized as
\be
\delta p = \cs \rho \delta + \frac{3 a H(\cs - \ca)}{k^2} \rho V ,
\ee
where the adiabatic speed of sound $\ca \equiv \dot{p}/\dot{\rho} = w$ for a fluid with constant equation of state.
Since we focus on late cosmological times, we can approximate $H$ with 
\be
H^2 = H_0^2\left[ \Omega_{m,0} a^{-3}+ (1-\Omega_{m,0} )a^{-3(1+w)} \right] ,
\ee
where $\Omega_{m,0}$ indicates the dark matter density parameter and $H_0$ is the Hubble parameter today.

To form a complete set of differential equations we add the Poisson equation, derived from the first and second Einstein equations,
\bea
k^2\phi  &=& -4\pi G a^2\sum_i\rho_i\left( \delta_i+\frac{3aH}{k^2}V_i\right) =\nonumber \\
&=& -4\pi G a^2\sum_i\rho_i\Delta_i \,, \label{eq.Poisson_general}
\eea
(where $\Delta_i \equiv \delta_i + 3 a H V_i/k^2$ is the gauge-invariant density perturbation of the 
$i-th$ fluid, the sum runs over all clustering fluids and $G$ is the Newton constant)
and the fourth Einstein equation,
\bea  \label{eq.4thEinsteinB}
k^2\left(\phi -\psi \right) &=& 12 \pi G a^2 \, (1+w) \rho\, \sigma  \\
&=& \frac{9}{2}  H_0^2 (1-\Om)a^{-(1+3 w)} (1+w)  \sigma  \nonumber \\
&\equiv& B(a)\,\sigma \,. \label{eq.4thEinstein}
\eea

\subsection{Analytical solutions for dark energy perturbations}

In the aforementioned approximations and assuming matter domination we have found in 
\cite{fingerprinting3} the following analytical solutions for $\delta$, $V$ and $\sigma$:
\bea
\delta &=& \frac{3(1+w)^2}{3 \cs (1+w) + 8 \left( \cs -w \right) \cv} \frac{\phi_0}{k^2}  ,  \label{eq:delta-sub-below}\\
V &=& - \frac{9(1+w)^2 \left( \cs -w \right)}{3 \cs (1+w) + 8 \cv (\cs -w)} H_0 \sqrt{\Omega_m} \frac{\phi_0}{\sqrt{a}k^2} , \nonumber \\
&=& -3  a H \left(\cs -w \right) \delta , \label{eq:V-sub-below}\\
\sigma &=& - \frac{8 \cv \left( \cs -w\right)}{3 \cs(1+w) + 8 (\cs -w) \cv} \frac{\phi_0}{k^2} ,
\label{eq:sigma-sub}
\eea 
where the constant $\phi_0$ is defined from the relation $k^2 \phi \simeq -\phi_0$, 
which is valid strictly only during matter domination and while neglecting dark energy perturbations.

As already shown in \cite{fingerprinting3}, modes above the sound horizon at early times 
are effectively uncoupled from the anisotropic stress, since the term on the right hand side of 
Eq. (\ref{eq:sig}) is small compared to the terms on the left hand side. Here dark energy 
perturbations follow the standard evolution for $\cv = 0$, which was found in \cite{fingerprinting1}:
\bea
\delta &=& (1+w)\frac{\phi_0}{\cs k^2} , \label{eq:delta-std} \\
V &=& - \frac{3 (1+w) \left(\cs -w \right) H_0 \sqrt{\Om}}{\cs k^2} a^{-1/2} ,  \\
\sigma &=& 0.
\eea
We remark that, as can be seen by comparing Eqs. (\ref{eq:delta-sub-below}) 
and (\ref{eq:delta-std}), the damping introduced by viscosity is stronger by a 
factor $(1+w)$ with respect to the standard case of isotropic dark energy.
The Eqs.~(\ref{eq:delta-sub-below})-(\ref{eq:sigma-sub}) can be rewritten in terms of an 
effective sound speed \cite{fingerprinting3}
\be
\ce= \cs+\frac{8}{3}\cv\frac{\cs-w}{1+w}\,.
\label{eq:ceff}
\ee
This means that the important quantity determining the growth of the dark energy perturbations 
is a combination of the sound speed and the viscosity. These have a similar damping effect on 
density and velocity perturbations. 

We also point out that while normally the case $w=-1$ represents a singularity for dark energy perturbations, 
the situation is less clear when anisotropic stress is present. In general it is enough to keep $(1+w)\sigma$ finite
as $w\rightarrow -1$. In addition, although the source term of Eq.\ (\ref{eq:sig}) appears singular in this limit,
we can see from Eq.\ (\ref{eq:V-sub-below}) that $V$ decays $\propto (1+w)^2$ for our model and so the term does not
actually diverge.

\section{Observable effects of viscosity on galaxy clustering and weak lensing} \label{sec:derivatives}

Once we have shown the analytic expression of density and velocity perturbations and of the anisotropic stress, let us see how these enter our observables. In particular, 
in this section we evaluate the impact of viscosity 
on galaxy clustering and weak lensing maps, in order to understand what to expect from the error forecasts.  
To do this we look at the derivatives of our observables 
with respect to the parameters $\cs$ and $\cv$. These derivatives will appear in the Fisher matrix computation: 
the larger the derivative, the stronger the dependence of our observable  from the analyzed parameter, 
and the smaller the forecasted error.
 We analyze separately each different component characterizing these observables and evaluate analytically 
its impact in the total derivative, with the help of the analytical parameters introduced in \cite{deparameters}: 
the clustering parameter $Q$ and the anisotropy parameter $\eta$. 

\subsection{Parameterizing dark energy perturbations}

First of all, we shortly describe our parameters, which will help us understanding the behavior of our 
observables in the presence of a non-null $\cv$. The clustering parameter $Q \equiv 1+ \rho\Delta/(\rho_m\Delta_m)$ (where $\Delta$, $\Delta_m$ are gauge-invariant
comoving density perturbations),
quantifying the size of the dark energy perturbations compared to the matter perturbations (and hence to 
the total perturbations, given that dark energy perturbations are much smaller than matter ones), parameterizes the 
deviation from a purely matter-dominated Newtonian potential.
In the presence of viscosity we have obtained \cite{fingerprinting3}
\bea
Q -1 &=& \frac{1-\Om}{\Om}(1+w)\frac{a^{-3w}}{1-3w + \frac{2 k^2 a}{3H_0^2\Om}\ceff} \nonumber \\
&=& Q_0\frac{a^{-3w}}{1+\alpha\,a},
\label{eq:qtot}
\eea
where $\alpha =  2 k^2 \ce/[(3 H_0^2 \Om)(1-3w)]$, $Q_0 = (1+w)(1-\Om)/\left[\Om (1-3w)\right]$ and 
$\ce$ is given by Eq.~(\ref{eq:ceff}).
Notice that, as explained in \cite{fingerprinting3}, Eq. (\ref{eq:qtot}) is valid also for $\cs = \cv = 0$.

As regards the anisotropy parameter, this is defined as 
\be \label{eq:eta-def}
\eta \equiv \frac{\psi}{\phi} -1.
\ee
It is zero in the case of standard General Relativity with non-anisotropic fluids because then the two metric perturbations $\phi$ and $\psi$ are equal. When anisotropic stress is present instead, $\phi$ and $\psi$ are different. In our case $\eta$ is indeed non-null and given by
\be
\eta = -\frac{9}{2}H_0^2(1-\Om)(1+w)\frac{a^{-1-3w}}{k^2 Q}\left(1- \frac{\cs}{\ceff} \right)\,. \label{eq:etatot}
\ee
Having established the dependence of $Q$ and $\eta$ on $\cs$ and $\cv$, we evaluate then how the main ``ingredients'' of WL power spectrum and galaxy power spectrum, i.e. the matter power spectrum, the redshift space distortions and the weak lensing potential  depend on $Q$ and $\eta$, hence on the speed of sound and the viscosity parameter.

\subsection{The dark matter power spectrum\label{sec:pk}}

\begin{figure}
\centering
\epsfig{figure=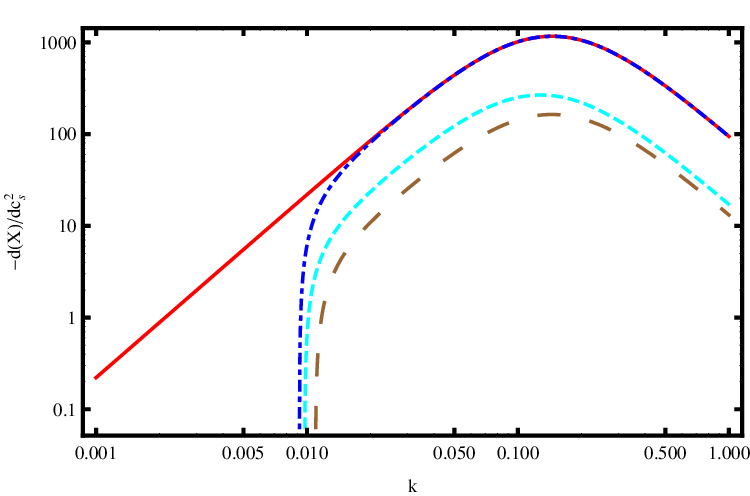,width=3.4in}
\epsfig{figure=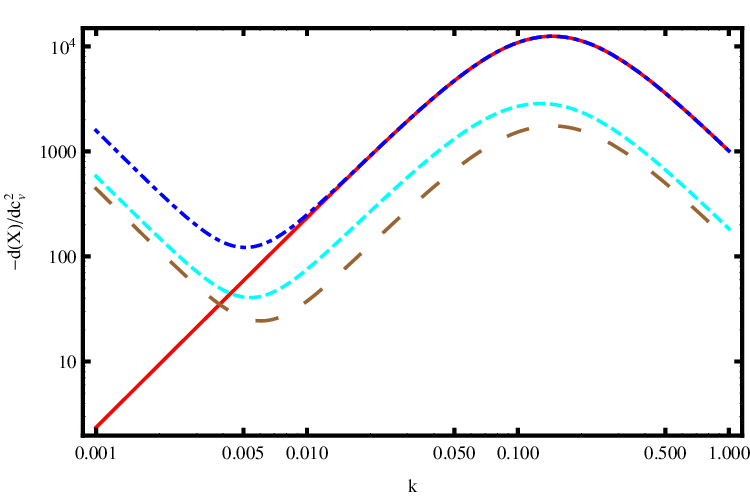,width=3.4in}
\caption{Here we show the sensitivity of the parameters entering the WL and GC observables to the sound speed $\cs$ and the anisotropy parameter $\cv$. Plotted are the derivatives of $Q(a,k)$ (red solid lines), $\Sigma(a,k)$ (blue dotted-dashed lines), $G(a,k)$ (cyan short-dashed lines),  and $f(a,k)$ (brown long-dashed lines) with respect to $\cs$ (top panel) and to $\cv$ (bottom panel), as a function of the wavenumber $k$ in units of $h/$Mpc.
The viscosity term is set to $\cv=10^{-6}$ and the sound speed to $\cs = 10^{-6}$. We can see that the most sensitive parameters are $\Sigma$ and $Q$ and that these two parameters are degenerate (i.e. have the same amplitude and shape) if $k$ is larger than $\sim 0.1 h/$Mpc.}\label{fig:dXdcscv}
\end{figure}

\begin{figure*}
\centering
\epsfig{figure=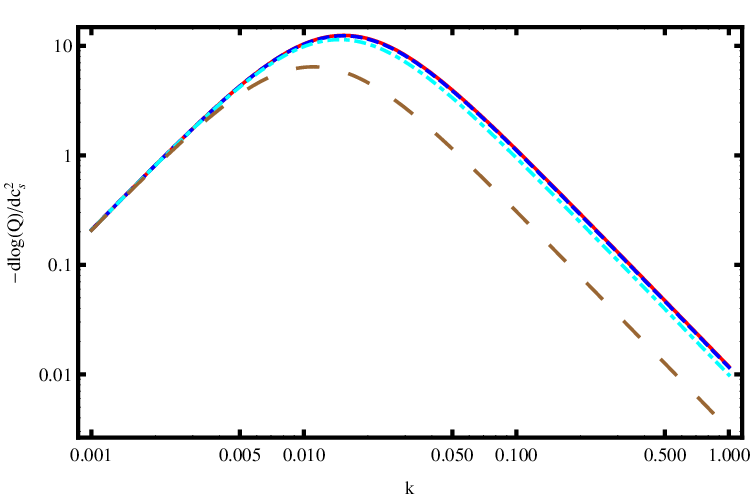,width=3.4in}\quad
\epsfig{figure=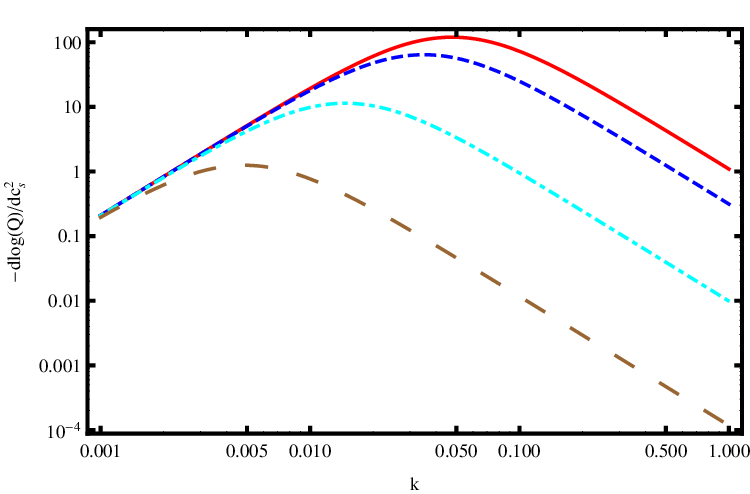,width=3.4in}
\vspace{0.1in}
\epsfig{figure=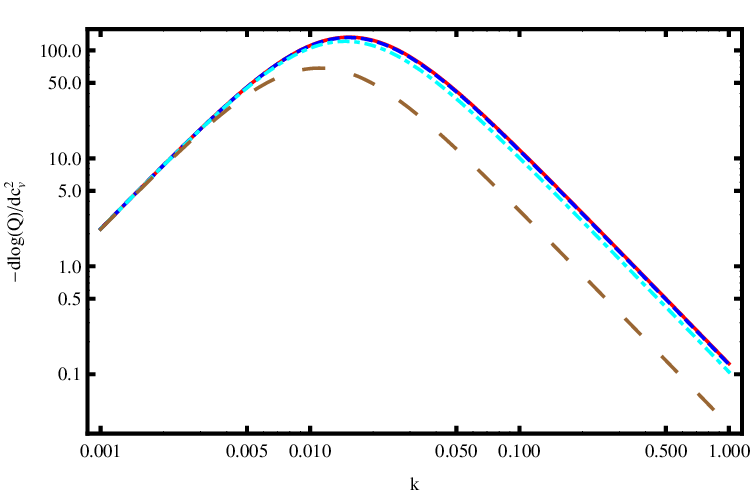,width=3.4in}\quad
\epsfig{figure=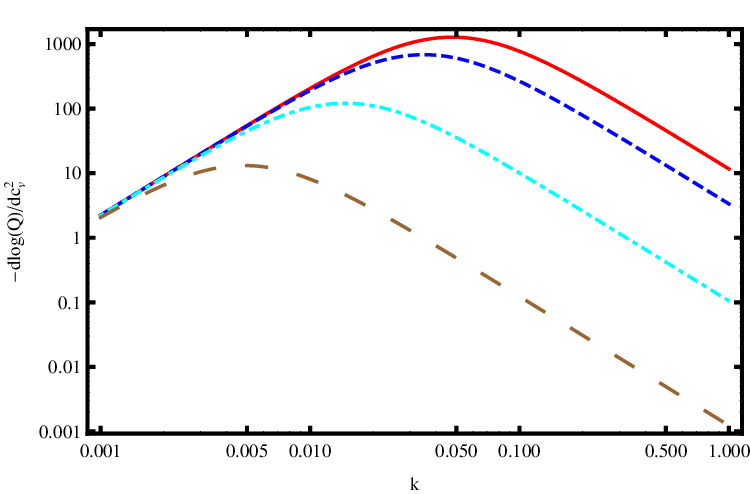,width=3.4in}
\caption{Here we show how the sensitivity of $Q$ to the sound speed $\cs$ and the viscosity parameter $\cv$ depend on the amplitude of $\cs$ (left panels) and $\cv$ (right panel). We plot the derivative of  the logarithm of $Q(a,k)$ with respect to the sound speed $\cs$ (top panels) and to the viscosity parameter $\cv$ (bottom panels) 
as a function of the wavenumber $k$ in units of $h/$Mpc. 
On the left panels, we fix the viscosity term $\cv=10^{-6}$ and the sound speed $\cs$ takes values $10^{-3}, 
10^{-4}, 10^{-5}$ and $10^{-6}$, brown long dashed, cyan dot dashed, blue dashed and red solid line, respectively.
On the right panels, we fix the viscosity term $\cs=10^{-6}$ and the sound speed $\cv$ takes values $10^{-3}, 
10^{-4}, 10^{-5}$ and $10^{-6}$, brown long dashed, cyan dot dashed, blue dashed and red solid line, respectively.
It can be seen that reducing the value of $\cs$ or $\cv$ improves the sensitivity of $Q$ to them, but the smaller $\cs$ or $\cv$, the smaller the improvement. We also see that $Q$ is a factor of $\sim 10$ more sensitive to $\cv$ than $\cs$ (compare corresponding top and bottom panels). This is because $Q$ depends on sound speed and viscosity through $\ce$, in whose expression a factor of $\sim 10$ multiplies $\cv$ while $\cs$ is multiplied by a factor of $1$.
}\label{fig:dQ}
\end{figure*}

The linear matter power spectrum $P_m(k, a)$ can be expressed as the product of today's $P_m(k)$ 
and its redshift evolution $G(a,k)$. 
Today's matter power spectrum $P_m(k)$ is affected by dark energy perturbations, hence by $\cs$ and $\cv$, through a linear dependence on $Q$. This is because the matter density contrast $\delta_m$ is sourced by the gravitational potential $\phi$, which is in turn modified by the presence of $\delta$. The Poisson equation (\ref{eq.Poisson_general}) can indeed be expressed as
\be
k^2 \phi = -4 \pi G a^2 Q \rho_m \Delta_m .
\label{eq:phi-Q}
\ee
Hence, we can try to assess how sensitive 
the matter power spectrum today is to changes of the sound speed and the viscosity, by computing the derivatives of $Q$ with respect to $\cs$ and $\cv$:
\bea
\frac{\partial Q}{\partial  \cs} &=&  -\frac{\alpha a}{1+\alpha a} (Q-1)  \frac{\ce -w}{\ce(\cs -w)} , \\
\frac{\partial  Q}{\partial  \cv} &=&  -\frac{\alpha a}{1+\alpha a} (Q-1) \frac{\ce -\cs}{\ce\cv}. \label{eq:dQdcv}
\eea
The derivatives of $Q$ with respect to $\cs$ and $\cv$ are shown
in Fig. \ref{fig:dXdcscv} (red solid lines).
In Fig. \ref{fig:dQ} we then show how the derivatives of $Q$ with respect to $\cs$ vary when changing the fiducial model: 
on the left panels we vary the fiducial $\cs$, which takes the values $\cs = 10^{-6}$, $10^{-5}$ and $10^{-4}$ and $\cv$ is 
fixed to $10^{-4}$, while on the right panels $\cv = 10^{-6}$, $10^{-5}$ and $10^{-4}$ and $\cs = 10^{-4}$
(we always use units where the speed of light is $c=1$). 
Our first consideration looking at the figure is that reducing the fiducial value of $\cs$ or $\cv$ improves the 
sensitivity of $Q$ to them, but not indefinitely: the smaller $\cs$ or $\cv$, the smaller the improvement. 
Secondly, we notice that $Q$ is a factor of $\sim 10$ more sensitive to $\cv$ than $\cs$ (compare corresponding top and bottom panels). 
The reason for this is to be found in the dependence of $Q$ on $\cs$ and $\cv$ through $\ce$: 
in its expression a factor of $\sim 10$ multiplies $\cv$, so that the sensitivity of $\ce$ to $\cv$ is 
better by such factor than that to $\cs$.

\subsection{The growth factor}

The growth factor, characterizing the change of the matter power spectrum with respect to today's shape and amplitude, is given by
\be
G(k, a) = \int_{a_0}^a \frac{f(a',k)}{a'} da'
\ee
where $a_0$ is today's scale factor and the growth rate $f$ can be expressed as
\be \label{eq:fdef}
f(a, k) = \Omega_m(z)^{\gamma(k,z)}
\ee
 and the growth index $\gamma (k,z)$ can be written in terms of $Q$ and $\eta$ \cite{gammapar}:
\be
\gamma=\frac{3\left(1-w-A\left(Q,\eta\right)\right)}{5-6w} ,
\label{eq:gamma-Qeta}
\ee
with 
\be
A\left(Q,\eta\right)=\frac{\left(1+\eta\right)Q-1}{1-\Omega_{m}\left(a\right)} .
\label{eq:A-Qeta}
\ee
using Eqs. (\ref{eq:gamma-Qeta}-\ref{eq:A-Qeta}) we can compute the derivatives 
$\partial  G / \partial  \cs$, $\partial  G / \partial  \cv$:
\bea
\frac{\partial  G}{\partial  \cs} &=& - G \frac{3 Q_0}{5-6 w} \frac{\ce -\cs}{\ce (\cs -w)} \int_{a_0}^
{a}  \left[ \frac{3w}{\alpha} x^{-3 w -2} + \right.\nonumber \\
&+&\left. \alpha \frac{\ce - w}{\ce -\cs} \frac{x^{-3 w}}{(1+\alpha x)^2} \right] dx , \\
\frac{\partial  G}{\partial  \cv} &=& - G \frac{3 Q_0}{5-6 w} \frac{\ce -\cs}{\ce\cv} \int_{a_0}^
{a}  \left[ \frac{3 \cs}{\alpha} x^{-3 w -2} + \right.\nonumber \\
&+&\left. \alpha \frac{x^{-3 w}}{(1+\alpha x)^2} \right] dx \,.
\eea
These are shown in Fig. \ref{fig:dXdcscv} (cyan short-dashed line). Comparing the derivatives with respect to $\cs$ and $\cv$ of $G$ with those of $Q$ we notice that the former are smaller than the latter, implying a smaller sensitivity of the latter to the sound speed and the viscosity. As also explained in \cite{fingerprinting2}, this is because $G$ depends on the integral of $Q$:  the growth factor is not probing the deviation of Q from unity, but rather how Q evolves with time.

\subsection{The weak lensing potential}\label{sec:wlpot}

\begin{figure}
\centering
\epsfig{figure=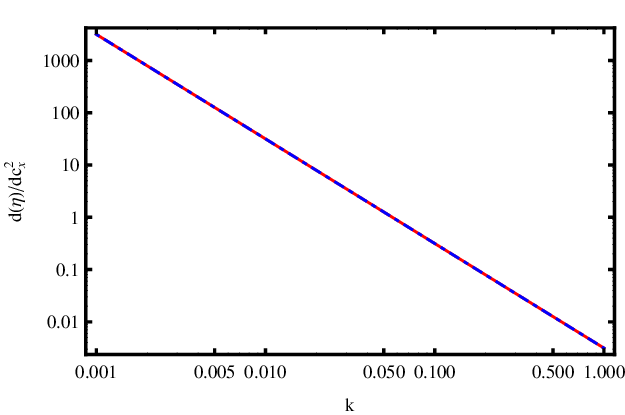,width=3.4in}
\caption{Here we show the sensitivity of the anisotropy parameter $\eta$ to the sound speed $\cs$ and viscosity parameter $\cv$. Derivatives of $\eta(a,k)$ with respect to the sound speed $\cs$ (solid red line) and the viscosity parameter $\cv$ (blue dotted-dashed line)  are shown
as a function of the wavenumber $k$ in units of $h/$Mpc. 
Here the fiducial viscosity term is $\cv=10^{-6}$ and the fiducial sound speed is $\cs = 10^{-6}$. By comparing the size of these derivatives to that of the corresponding derivatives of $Q$ of Fig. \ref{fig:dXdcscv} we notice that  the first are much smaller than the second when $k>0.01 h/$Mpc, so that their contribution to the derivatives of $\Sigma$ (related to the WL potential) is very small at smaller scales.
}\label{fig:deta}
\end{figure}

In WL experiments, the important quantity is the lensing potential $\Phi = \phi+\psi$.
As regards $\phi$, the Poisson equation couples it to the dark energy contrast, and we expressed the influence of the latter through the parameter $Q$, see Eq. (\ref{eq:phi-Q}).
The potential $\psi$ instead is related to the anisotropic stress, and the parameter connecting it to $\phi$ is $\eta$, as in Eq. (\ref{eq:eta-def}).
Hence, using Eqs. (\ref{eq:phi-Q}) and (\ref{eq:eta-def}) we find the following for $\Phi$:
\be
k^2\Phi = -8\pi\,G\,a^2\left(1+\frac{1}{2}\eta\right)Q\,\rho_m\Delta_m\,.
\label{eq:phi-wl}
\ee
The resulting quantity is therefore a combination of the anisotropic stress and the 
dark energy density contrast: 
\be
\Sigma = \left(1+\frac{1}{2}\eta\right)Q\,,
\label{eq:sigma-wl}
\ee
and it represents the deviation of the WL potential from the standard case of no dark energy perturbations.
We can compute now the derivatives of $\Sigma$ with respect to the sound speed $\cs$ and the viscosity term $\cv$: 
\bea \label{eq:dSigmadcs}
\frac{\partial\Sigma}{\partial\cs} &=&-\left(Q-1\right)\left\{\frac{\alpha\,a}{1+\alpha\,a}\frac{1}{\ce}\left[1+\frac{8}{3}\frac{\cv}{1+w}\right] + \right.\nonumber \\
&+&\left. 4\frac{1+\alpha\,a}{\alpha\,a}\frac{\cv}{\ce}\frac{w}{1+w}\right\}\\
\frac{\partial\Sigma}{\partial\cv} &=&-\left(Q-1\right)4\frac{\cs-w}{1+w}\left\{\frac{\alpha\,a}{1+\alpha\,a}\frac{1}{\ce}\frac{2}{3}+ \right.\nonumber \\
&+&\left.\frac{1+\alpha\,a}{\alpha\,a}\frac{\cs}{\ce}\right\}\,, \label{eq:dSigmadcv}
\eea
and plot them in Fig. \ref{fig:dXdcscv} together with the other derivatives. We clearly see that at scales smaller than $k \sim 0.01 h/$Mpc  the derivatives with respect to $\cs$ or $\cv$ of $\Sigma$ have the same amplitude and shape as those of $Q$. Looking at the expression of $\partial\Sigma/\partial \cs$, $\partial\Sigma/\partial \cv$, Eqs. (\ref{eq:dSigmadcs}-\ref{eq:dSigmadcv}), this means that $\eta$ only plays a role at very large scales, while its contribution is very small at all other scales. We can also see that directly in the expression for $\eta$,
Eq.\ (\ref{eq:etatot}), which contains a factor $(Ha/k)^2$ relative to $Q-1$, Eq.\ (\ref{eq:qtot}), and so is suppressed on sub-horizon scales.
A confirmation of this can be found in Fig. \ref{fig:deta}  which shows the evolution with $k$ of the derivatives of $\eta$, which are much smaller than those of $Q$. 
We also notice, comparing upper to lower panel of Fig. \ref{fig:dXdcscv}  that for small values of the speed of sound and the viscosity (as those selected to produce the plot), the derivatives with respect to $\cs$ are identical in shape to those with respect to $\cv$, if we consider smaller scales.

\subsection{Redshift space distortions\label{sec:zdist}}

Galaxy redshift surveys do not directly observe the total matter distribution: they produce a 3D map of the 
galaxy distribution, where the information on the radial distance to  each galaxy is obtained through the measured galaxy redshift. 
Due to the galaxy peculiar velocities, the redshift map is distorted with respect to the real space map. Such redshift space distortion (RSD) was first 
described by \cite{Kaiser}, who modeled it through a factor $(1+ (f (k, z)/b) \mu^2)^2$ multiplying the 
matter power spectrum, where $f$ is the growth rate
 (see Eq. (\ref{eq:fdef})),  $b$ is the bias factor relating the amplitude of the galaxy power spectrum to the matter 
power spectrum, and $\mu$ is the cosine of the  component of $k$ parallel to the line of sight.

For this reason, we evaluate here the sensitivity of $f$ to sound speed and viscosity parameter by computing   the derivatives of $f$  with respect to $\cs$ and $\cv$. 
These are given by $\partial f/\partial \cs = f \ln \Omega_m(a) \partial \gamma/\partial \cs$ (and equivalently for $\cv$). 
Since the main dependence on viscosity and speed of sound will come from the term 
$\partial \gamma /\partial \cs$ (given that the term $f$ will be close to $1$ during matter domination), 
we only show here the analytical derivatives of $\gamma$:
\bea
\frac{\partial \gamma}{\partial \cs} &=& \frac{3 (Q-1)}{(5-6w) (1-\Omega_m(a))} \frac{\ce -\cs}{\ce (\cs -w)} \times\nonumber\\
&\times& \left[ 3 w \frac{1+\alpha a}{\alpha a} + \frac{\ce - w}{\ce - \cs} \frac{\alpha a}{1+ \alpha a} \right] , \\
\frac{\partial \gamma}{\partial \cv} &=& \frac{3 (Q-1)}{(5-6w) (1-\Omega_m(a))} \frac{\ce -\cs}{\ce \cv} \times\nonumber\\
&\times&\left[ 3 \cs \frac{1+\alpha a}{\alpha a} + \frac{\alpha a}{1+ \alpha a} \right] .
\eea
The logarithm of the derivatives of the Kaiser term with respect to $\cs$ and $\cv$ (integrated over the angle $\mu$) are shown as well in Fig. \ref{fig:dXdcscv}. Their shape is very similar to that of the derivatives of $G$ and of $\Sigma$ but they have always smaller size than them, which means that the contribution of the RSD in the detection of $\cs$ and $\cv$ should not be very important.

\subsection{Comparison of galaxy power spectrum and weak lensing potential \label{sec:comp}}

We want now to identify exactly the quantities entering the galaxy power spectrum and the WL power spectrum, in order to understand which of them is more sensitive to the sound speed and the viscosity parameter. 

Let us start with the former: it will depend on the matter power spectrum and on the RSD factor, hence the relevant parameters for detecting DE perturbations will be $Q$ (see Eq. (\ref{eq:phi-Q})) and the growth factor $G$, related to the matter power spectrum today and its evolution with time, respectively, and the growth rate $f$, detectable through the RSD term.
If we consider the WL power spectrum, this will again depend on the matter power spectrum, hence on $Q$ and $G$, and also on the difference between the two gravitational potentials, proportional to $\Sigma$, as from Eq. (\ref{eq:phi-wl}).
We see from Fig. \ref{fig:dXdcscv} that the contribution from $f$ is the smallest, and that the sensitivity to variations in $\cs$ and $\cv$ of $G$ is slightly lower than that of $\Sigma$ and $Q$.
This means that in principle WL surveys, measuring $\Sigma$, should be slightly more sensitive than galaxy redshift surveys, which measure $f$.\footnote{
To be precise, as pointed out e.g. in \cite{Amendola:2012ky}, the quantities which can really be observed are the combinations $f/b$ and $f + f'/f$ and not $f$ itself. Since the sensitivity of $f$ to $\cs$ and $\cv$ is very small, and since $b$ is of order unity, this does not change our conclusions.}
However, other considerations have to be made. 
Derivatives of $\Sigma$ and $Q$ differ in shape and amplitude only in the region of very large scales. As this region is not  probed by WL measurements (because of the size of the survey and of the projection on the sphere), constraints to $\cs$ and $\cv$ will be degenerate: it will be very difficult to separately measure them.

Moreover, the actual result will depend also importantly on the properties of the single surveys, ``weighted'' by the sensitivities discussed above.

\section{Forecasts for the Euclid survey} \label{sec:Fisher-forecasts}

With the help of the sensitivity of our observables to $\cs$ and $\cv$ that have evaluated analytically,
we can now forecast the actual precision with which the Euclid WL and GC surveys will be able to
measure these parameters.
As already mentioned, Euclid \cite{RedBook} is a medium-size mission of the ESA Cosmic
Vision programme,  recently adopted for implementation, whose launch is planned for 2020. It will perform two surveys: a photometric survey in the visible and in three
near-infrared bands, to measure weak gravitational lensing maps by imaging $\sim 1.5$ billion galaxies, and a spectroscopic slitless
survey of $\sim 65,000,000$ galaxies. Both surveys will be able to constrain
both the expansion and growth history of the universe and  will cover a total area of $15,000$ square deg.

Our fiducial Euclid survey follows the specifications which can be found in the Euclid Definition 
Study Report (also called Red Book) \cite{RedBook}, and correspond to the most up-to-date simulations of Euclid's performance.

As a fiducial model for our Fisher analysis we choose the WMAP-7 flat $\Lambda$CDM cosmology,
as also used in the Euclid  Red Book \cite{RedBook}, 
with the exception of the value of $w$, which we set to $w=-0.8$. This means that we have 
$\Om h^2 = 0.13$, $\Omega_{b,0} h^2 = 0.0226$, $\Omega_\Lambda = 1-\Om =  0.73$, 
$H_0 = 71 $, $n_s = 0.96$ (where $n_s$ is the scalar spectral index). 
The matter power spectrum was computed using CAMB\footnote{http://camb.info} \cite{camb}.

\subsection{Weak lensing}

We start by investigating the sensitivity of the Euclid WL survey to the dark energy parameters. 
We proceed as in \cite{deparameters} and \cite{fingerprinting2}, the only difference here being that we add the contribution 
of the anisotropic stress. 

Following Eqs.~(\ref{eq:phi-wl}-\ref{eq:sigma-wl}) we can evaluate the convergence WL power spectrum 
(which in the linear regime is equal to the ellipticity power spectrum): this is a linear function of the matter power spectrum convoluted 
with the lensing properties of space. For a $\Lambda$CDM cosmology it can be written as \bea
P_{ij}(\ell) &=& H_0^4\int_{0}^{\infty}\frac{{\rm d}z}{H(z)}W_{i}\lzr W_{j} \lzr \times \nonumber \\ && \times \,\, P_{nl}\left[P_{l}\left(\frac{H_0\ell}{r\lzr}, z\right)\right]\,,
\eea
where $\ell$ is the multipole number, $W_{i}$'s are the window functions, $P_{nl}\left[P_{l}\left(k, z\right)\right]$ is the non-linear power spectrum 
at redshift $z$ obtained correcting the linear matter power spectrum $P_{l}\left(k, z\right)$, see \cite{deparameters} for more details.

When dark energy perturbations come into play, the former gets modified. 
There is no easy way to modify the convergence power spectrum when we consider the non-linear scales,
 because we need to evaluate the WL power spectrum, i.e. $\langle\Phi^2\rangle$, 
which only at linear scales is simply proportional to $\Sigma^2\langle\Delta_{m}^2\rangle$. 
If we consider non-linear scales, the above expression is no longer strictly valid because 
$\Sigma$ is a first order quantity.  In practice, what we would need to do is to convolute 
the modified linear matter power spectrum $P_{nl}\left(\Sigma^2\,P(k)\right)$. 
We instead compute $\Sigma^2\,P_{nl}\left(P(k)\right)$ 
\footnote{The reason for this is that to include the effect of dark energy perturbations in the non-linear power spectrum we cannot make use of the analytic expression of $P_{nl}$, which is designed for the case of absence of dark energy perturbations. So if we want to take the latter into account we need to solve the Boltzmann equations, hence use CAMB, which does not, however, compute $P_{nl}(\Sigma^2 P_l)$ but only $P_{nl}(P_l)$. Therefore we decided to make the aforementioned approximation: $P_{nl}(\Sigma^2 P_l) \sim \Sigma^2 P_{nl}(P_l)$, although it is unclear how large is the error we commit when making it.}. 

The {\em modified} convergence power spectrum is then
\bea
P_{ij}(\ell) &=& H_0^4\int_{0}^{\infty}\frac{{\rm d}z}{H(z)}W_{i}\lzr W_{j}\lzr \times  \nonumber \\ 
&& \times \,\, \Sigma^2\,P_{nl}\left[P_{l}\left(\frac{H_0\ell}{r\lzr}, z\right)\right]\,.
\eea
The Fisher matrix for WL is then given by
\bea
F_{\alpha\beta} &=& f_{\rm sky}\sum_{\ell} \frac{\left(2\ell+1\right)\Delta\ell}{2} \times \nonumber \\
&& \times \,\, \partial\left(P_{ij}\right)_{,\alpha}C_{jk}^{-1}\partial\left(P_{km}\right)_{,\beta}C_{mi}^{-1} \,, \label{eq:FisherWL}
\eea
where $f_{\rm sky}$ is the observed fraction of the sky, the  partial derivatives represent $\partial /\partial\theta_{\alpha}$,  the corresponding cosmological parameters $\theta_{\alpha}$ 
are shown in Tab.~\ref{tab:cosmological-parameters-Pk} and
\be
C_{jk}= P_{jk}+\delta_{jk}\frac{\langle \gamma_{\rm int}^{1/2}\rangle}{n_j}\,,
\ee
where $\gamma_{\rm int}$ is the rms intrinsic shear (here we assume $\langle\gamma_{\rm int}^{1/2}\rangle$=0.22 \cite{amara}) 
and $n_{j}$ is the number of galaxies per steradians belonging to the $j$-th bin.

We compute the Fisher matrix  for 3 fiducial models: the case where $\cs = \cv = 10^{-6}$, 
$\cs = \cv = 10^{-4}$ and $\cs = 1$, $\cv = 0$. The remaining parameters which we allow to vary are 
here $\Om h^2$, $\Omega_b h^2$, $n_s$, $\Om$, $w_0$.

In the WL survey the redshift range covered is $0<z<2.5$, which we divide into 10 bins 
chosen such as to contain an approximately equal number of galaxies each.

\begin{figure}
\epsfig{figure=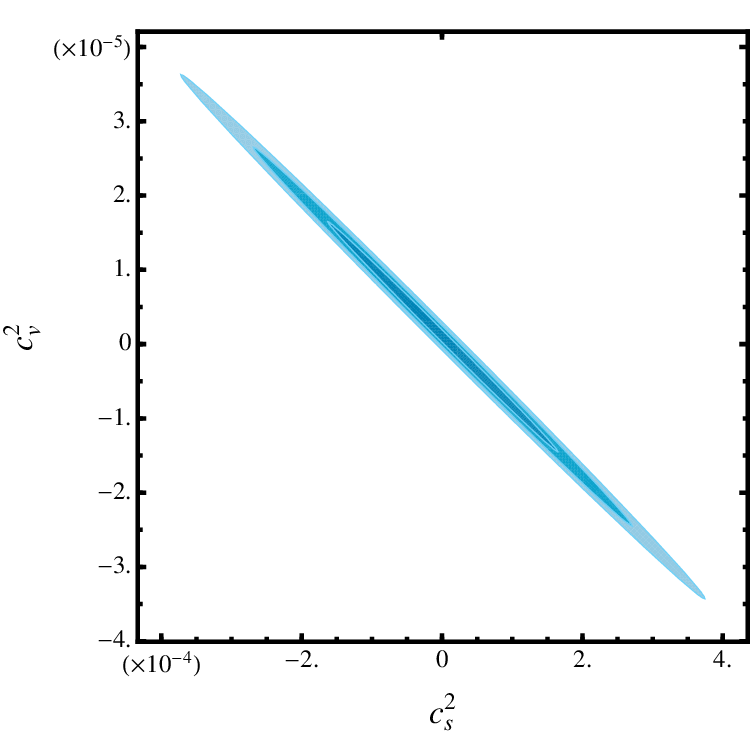,height=2.0in}
\epsfig{figure=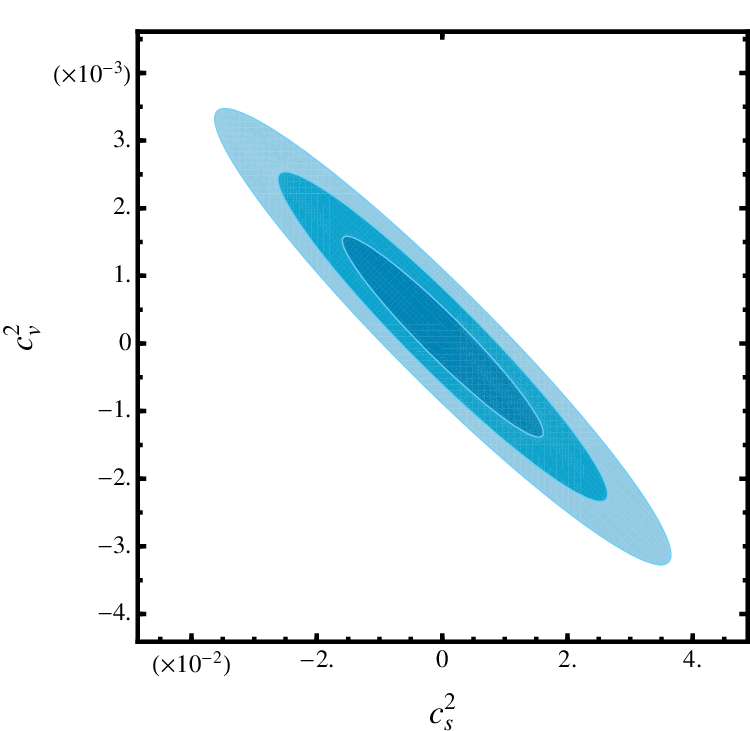,height=2.0in}
\epsfig{figure=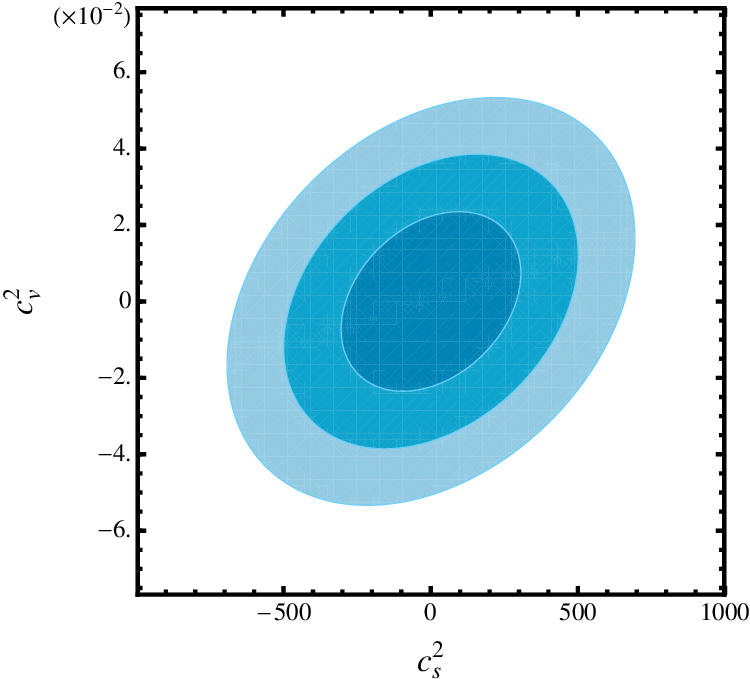,height=2.0in}
\caption{Here we plot the forecasted errors on sound speed $\cs$ and viscosity parameter $\cv$ from the Euclid WL survey. Shown are the 1,2 and 3 $\sigma$ Fisher ellipses for the parameters 
$\cs$ and $\cv$. 
The top panel corresponds to the fiducial $\cs = \cv = 10^{-6}$, the central panel to $\cs = \cv = 10^{-4}$ while the bottom one to the case $\cs = 1$ and $\cv = 10^{-6}$. In the case of the top and central panel, a very strong degeneracy between $\cs$ and $\cv$ can be seen. This is because when $\cs$ and $\cv$ are very small, the contributions from $\Sigma$ and $Q$ are very similar and both constrain $\ce$ rather than different combinations of $\cs$ and $\cv$.
}\label{fig:fish-wl-cs2cv210-6}
\end{figure}

Fig. \ref{fig:fish-wl-cs2cv210-6} shows the $1$, $2$ and $3\sigma$ Fisher ellipses for the parameters $\cs$ and $\cv$. As we can see, the $1\sigma$ errors are of the order of  $10,000\%$ on $\cs$ and of $1000 \%$ on $\cv$, but more importantly we immediately notice that there is a very strong degeneracy between $\cs$ and $\cv$.

To explain it, we should first remember that the parameters measured by WL are $Q$, $\Sigma$ and $G$ but as shown in Fig.  \ref{fig:dXdcscv}, the contribution 
of the growth factor $G$ to the total derivatives is small, hence the relevant parameters are $Q$ and $\Sigma$.
Let us look at the derivative of the weak 
lensing parameter $\Sigma$ with respect to the viscosity term $\cv$, 
ie. Eq.~(\ref{eq:dSigmadcv}), and rewrite it as: 
\bea
\frac{\partial\Sigma}{\partial\cv} &=&- Q_0 a^{-3w}\frac{\alpha\,a}{\left(1+\alpha\,a\right)^2}\frac{1}{\ce}\frac{8}{3}\frac{\cs-w}{1+w}  \nonumber \\&&+ 
\frac{1+\alpha\,a}{\alpha\,a}\frac{\cs}{\ce}4\frac{\cs-w}{1+w}\label{eq:dSigmadcv-1}\,,
\eea
where the first term on the right hand side is simply the derivative of $Q$ with respect to $\cv$ (see Eq.~(\ref{eq:dQdcv})), while the second term 
is directly connected to the anisotropic stress, hence to $\eta$. 
It can be seen that for small enough values of $\cs$ and $\cv$, $\alpha a \propto k^2 a$ is always smaller than $1$. In particular, when $k$ tends to zero the second term dominates and grows, as can be seen from Figs. \ref{fig:deta} and \ref{fig:dXdcscv}. Instead, when $k$ grows, and while $(1+\alpha a) \sim 1$, the first term behaves like $k^2$ and the second like $1/k^2$, so that the first term dominates.

Therefore, when the sound speed and the viscosity terms are small enough, for example when $\cs=\cv=10^{-6}$, then the first term on the right hand side of
Eq.~(\ref{eq:dSigmadcv-1}) is the dominant component - this is due to the relative increase of the dark energy perturbations, which is measured by $Q$.
When we increase the value of $\cs$, the first term in Eq. (\ref{eq:dSigmadcv-1}) starts 
to get smaller, as $\alpha a \sim 1$,  the two terms become comparable in size, and they contribute equally to the total derivative. 

We can show this numerically: for modes below the causal horizon with $k=200H_0$, and assuming $\cs=\cv=10^{-6}$,  
the first term in Eq.~(\ref{eq:dSigmadcv-1}) is of the order of $10^{4}$ whereas the second term is of  order  unity. 
If we set $\cs=1$ and $\cv=0$ instead, we have that  both terms become of the order of $10^{-4}$. 

So if both values of $\cs$ and $\cv$ are very small, the largest contribution in the derivative of $\Sigma$ comes from the term proportional to the derivative of $Q$. 
Since, as explained earlier, the two parameters measured by WL which contribute most strongly to 
the determination of $\cs$ and $\cv$ are $Q$ and $\Sigma$, and since their derivatives with 
respect to $\cs$ are proportional to those with respect to $\cv$ for small enough scales, 
given that here they carry \emph{almost} the same information, these parameters will 
be \emph{almost} degenerate. This is reflected by the top panel (and more mildly by the middle one) 
of Fig. \ref{fig:fish-wl-cs2cv210-6} and by our WL Fisher matrix, which will be \emph{almost} 
singular (see also \cite{Albrecht:2009ct} on this topic).

Moreover, we can understand the degeneracy by looking again at the first term on the right hand side of Eq. (\ref{eq:dSigmadcv-1}).
When $\cs$ is very small, the term $\cs-w \simeq -w$ so that all dependence on the speed of sound and viscosity parameter comes from $\ce$. The degeneracy direction will therefore be that of constant $\ce$, as can be easily verified from Fig. \ref{fig:fish-wl-cs2cv210-6}.
 An important consequence of this is 
that the inversion of the matrix will be unstable and the results will not be reliable, and the 
reason is that our data are not able to constrain the two parameters $\cs$ and $\cv$ at the 
same time but only a combination of them. This implies that the top panel of Fig. \ref{fig:fish-wl-cs2cv210-6} 
cannot be considered a reliable result apart from showing us the existence of a strong 
degeneracy between viscosity and speed of sound. This is also confirmed by comparing 
the elements of the relative Fisher matrix to the corresponding one for galaxy clustering. 
The middle panel is slightly closer to be reliable, while the bottom panel, where the 
fiducial values are $\cs =1$ and $\cv = 0$, is reliable because $\cs$ is large enough 
to break the degeneracy between $\cs$ and $\cv$.

\begin{table}
\begin{centering}\begin{tabular}{|c|c|c|c|}
\hline & \textbf{Parameters} & $\bf{P\left(k\right)}$ & \bf{WL} \tabularnewline
\hline 
\multicolumn{4}{|c|}{ }  \tabularnewline
\hline 
1& total matter density&  $\Omega_{m_{0}}h^{2}$ & $\Omega_{m_{0}}h^{2}$ \tabularnewline
\hline 
2 & total baryon density &$\Omega_{b_{0}}h^{2}$& $\Omega_{b_{0}}h^{2}$\tabularnewline
\hline 
3 & spectral index &  $n_{s}$ & $n_{s}$\tabularnewline
\hline 
4&matter density today&  $\Omega_{m_{0}}$& $\Omega_{m_{0}}$\tabularnewline
\hline 
5& equation of state parameter &$w_0$ & $w_0$ \tabularnewline
\hline 
6& sound speed & $\cs$ &  $\cs$\tabularnewline
\hline 
7& viscosity parameter & $\cv$ &  $\cv$\tabularnewline
\hline 
\multicolumn{4}{|c|}{ }  \tabularnewline
 \multicolumn{4}{|c|}{\bf{For each redshift bin}}  \tabularnewline
\multicolumn{4}{|c|}{ } \tabularnewline
\hline
8 & shot noise & $P_{s}$  & $\,$\tabularnewline
\hline 
\end{tabular}\par\end{centering}
\caption{Cosmological parameters (for the derivatives) for galaxy 
power spectrum and WL survey.
\label{tab:cosmological-parameters-Pk}}
\end{table}

We summarize our results in Table \ref{tab:errorsWL}, 
where we indicate the fully marginalized errors on $\cs$ and $\cv$. Let us recall again the caveat previously explained: errors associated to the fiducial models $\cs=\cv = 10^{-6}$ ($\cs = \cv = 10^{-4}$) are strongly (considerably) unreliable as they have been computed from an almost degenerate Fisher matrix. We list them here only for completeness.

\begin{figure}
\epsfig{figure=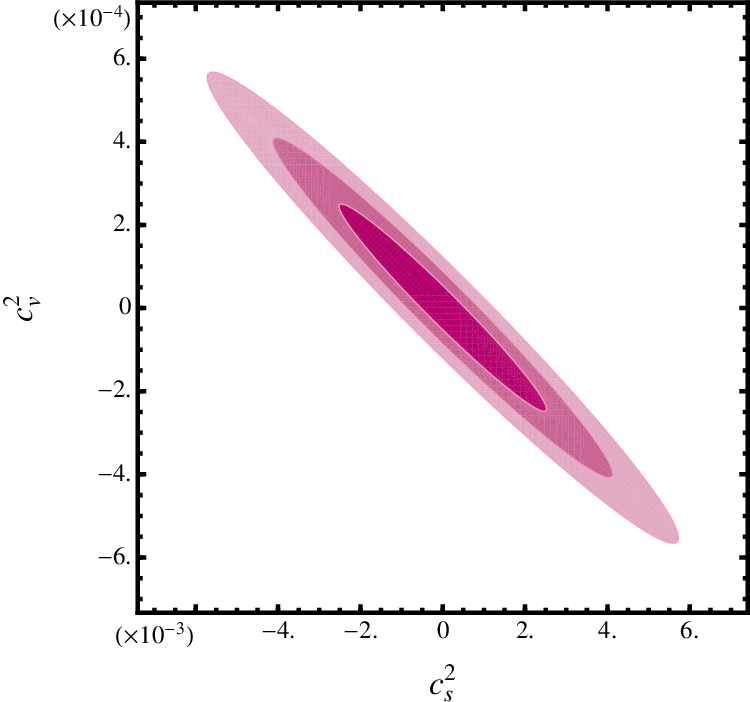,height=2.0in}
\epsfig{figure=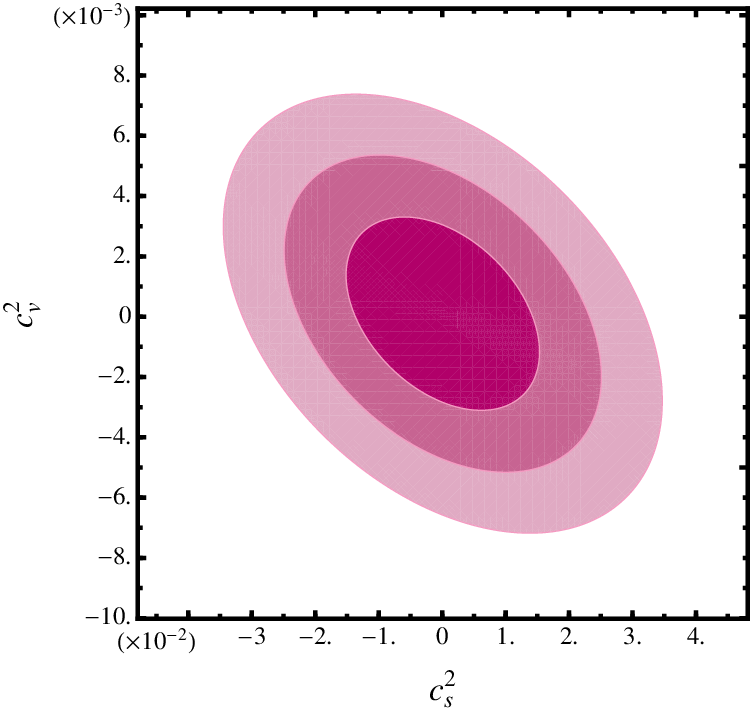,height=2.0in}
\epsfig{figure=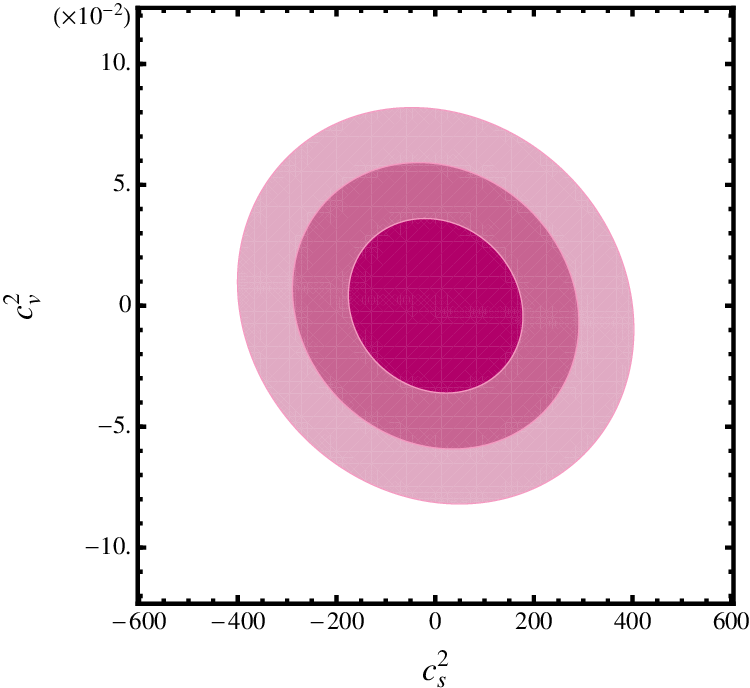,height=2.0in}
\caption{Same as Fig. \ref{fig:fish-wl-cs2cv210-6} but using only scales closer to linearity, with $\ell_{max} = 30$, which corresponds to $k_{max} = \,0.5 h/{\rm Mpc}^{-1}$. We notice that the degeneracy between $\cs$ and $\cv$ of Fig.  \ref{fig:fish-wl-cs2cv210-6} is removed, strongly for the cases of the two lowest panels. This is due to the fact that at larger scales the parameters $\Sigma$ and $Q$ start carrying information on both $\cs$ and $\cv$ and not only on a unique combination of them ($\ce$). The worsening of errors due to the use of a much smaller number of $\ell$s is not strong due to the fact that at smaller scales our observable is less sensitive to speed of sound and viscosity because here dark energy perturbations are suppressed.
}\label{fig:fish-wl-cs2cv210-6-linear}
\end{figure}

Before closing this section, let us look again at Eq. (\ref{eq:FisherWL}). The summation over $\ell$ has been stopped at the  default ``very optimistic'' $\ell_{max} = 5000$ used also in \cite{RedBook}, which falls into deeply non-linear scales (but regards the matter power spectrum only and is considered to be more reliable in the standard $\Lambda$CDM case). This amounts to believing  that we will have an adequate description of this regime for the anisotropic stress model by the time the Euclid satellite will be launched. Moreover, our results depend also on the assumptions made to compute the non-linear power spectrum (we use CAMB's implementation of the halo model and multiply it by $\Sigma$ to get the final power spectrum). We can now make two considerations: first, the non-linear $P(k)$ computed with the halo model is certainly not accurate at small scales; second $\Sigma$ is computed assuming linearity. Hence a more consistent approach would be to limit ourselves to less non-linear scales. We have therefore built alternative WL Fisher matrices using $k_{max} = 0.5\, h {\rm Mpc}^{-1}$, which corresponds to $\ell_{max} \simeq 30$ and falls less deeply into the non-linear regime. The results are shown in Fig. \ref{fig:fish-wl-cs2cv210-6-linear}. 
Here, apart from realizing that the forecasted errors become slightly larger (due to the fact that we use less information), we notice something interesting: the degeneracy has weakened considerably. The reason for this can probably be found by looking again at Fig. \ref{fig:dXdcscv}: at larger scales (corresponding to smaller $\ell$) the derivatives of $Q$  with respect to $\cs$ and $\cv$ start to differ from those of $\Sigma$. 
Hence, since for Fig. \ref{fig:dXdcscv} we used a smaller number of the $\ell$s  falling into the interval where $\partial \Sigma/\partial X \simeq \partial Q/ \partial X$, the total result is less degenerate.
In particular, in this case we can fully trust the results shown in the middle and lower panels of the aforementioned figure.
Looking at Fig. \ref{fig:dXdcscv} we can also understand another reason why removing so many $\ell$s from our Fisher matrix calculation does not make errors on $\cs$ and $\cv$ much larger, but they stay of the same order of magnitude (apart from the case $\cs = \cv = 10^{-6}$), compare Fig. \ref{fig:fish-wl-cs2cv210-6} with Fig. \ref{fig:fish-wl-cs2cv210-6-linear}. We see from Fig. \ref{fig:dXdcscv} that the scales most sensitive to $\cs$ and $\cv$ (i.e. those where the derivatives have largest amplitude) are those around $k \sim 0.01$. Below $k\sim 0.1$ there is little contribution, and this is due to the fact that dark energy perturbations are suppressed on small scales.

\begin{table}
\begin{centering}
\begin{tabular}{|c|c|c|c|}
\hline 
\multicolumn{4}{|c|}{WL}\tabularnewline
\hline 
\hspace{0.5cm}$\cs$\hspace{0.5cm}  & $\cv$  & $\sigma_{\cs}/\cs$ & $\sigma_{\cv}/\cv$ \tabularnewline
\hline 
$10^{-6}$  & $10^{-6}$  & $109 $  & $10.3 $ \tabularnewline
\hline 
$10^{-4}$  & $10^{-4}$  & $106$ & $9.81$ \tabularnewline
\hline 
$1$  & $0$  & $201$  & $ \sigma_{\cv} = 1.55 \times 10^{-2}$ \tabularnewline
\hline 
\end{tabular}
\par\end{centering}
\caption{Relative errors on the parameters $\cs$ and $\cv$ from the Euclid WL survey. For the case $\cv = 0$ the absolute error $\sigma_{\cv}$ is given.}
\label{tab:errorsWL} 
\end{table}

It is fair to make a final important remark: all the results we obtained depend on the parameterization we have used 
where the effects of the sound speed and the viscosity appear linearly in a unique expression Eq.~(\ref{eq:ceff}). 
This implies in particular that it may be possible that a different parameterization of the anisotropic 
stress removes the aforementioned degeneracy problem.

\subsection{Galaxy clustering}

Let us now show and comment the Fisher matrix forecasts for the
Euclid galaxy \emph{redshift} survey, computed for the same 3 fiducial models 
previously indicated for the WL case, i.e. $\cs=\cv = 10^{-6}$, $\cs=\cv = 10^{-4}$ and $\cs = 1$ $\cv =0$. 

Following \cite{seisen} we write the observed galaxy power spectrum as: 
\bea
P_{obs}(z,k_{r})  &=&  \frac{D_{Ar}^{2}(z)H(z)}{D_{A}^{2}(z)H_{r}(z)}
G^{2}(z)b(z)^{2}\left(1+\beta\mu^{2}\right)^{2}P_{0r}(k) \nonumber \\
&&  +  P_{shot}(z) \,,
 \eea
 where the subscript $r$ refers to the values at which we evaluate
the Fisher matrix, i.e. the reference (or fiducial) cosmological model. 
Here $P_{shot}$ is a scale-independent
offset due to imperfect removal of shot-noise, 
$\mu$ is the cosine of the angle of the wave mode with respect to the line of sight, $P_{0r}$ is the
 fiducial matter power spectrum evaluated at 
redshift zero, $G(z)$ is the linear growth factor of the matter 
perturbations, $b(z)$ is the bias
factor and $D_{A}(z)$ is the angular diameter
distance.
The wavenumber $k$ has also to be written in terms of the fiducial
cosmology (\cite{seisen} and see also \cite{aqg}
and \cite{sa} for more details).

The spectroscopic survey covers a redshift range of $0.65<z<2.05$,
 which we divide into 14 bins of equal 
width $\Delta z = 0.1$. As regards the bias, we assume it to be 
scale-independent, since this is a quite good approximation for
the large linear scales which we will use. Our fiducial bias was
derived by  \cite{Orsi}  using a semi-analytical model of
galaxy formation, and it is the same bias function used for the Euclid
 Red Book forecasts.  The expected galaxy number
densities which we used can be found in \cite{myeuclid} and 
were computed by using a sophisticated simulation \cite{Garillietal}. 
The scales $R$ used are such that $\sigma^2(R)\leq0.25$, with an 
additional cut at $k_{\rm max}= 0.20 \,h\,{\rm Mpc}^{-1}$ 
to avoid non-linearity problems. 

In the computation of the GC Fisher matrix we do not 
incur in the same degeneracy problem of the WL survey: 
our matrices are far from being singular, hence we can safely produce forecasts on the errors on $\cs$ and $\cv$.

\begin{figure}
\centering
\epsfig{figure=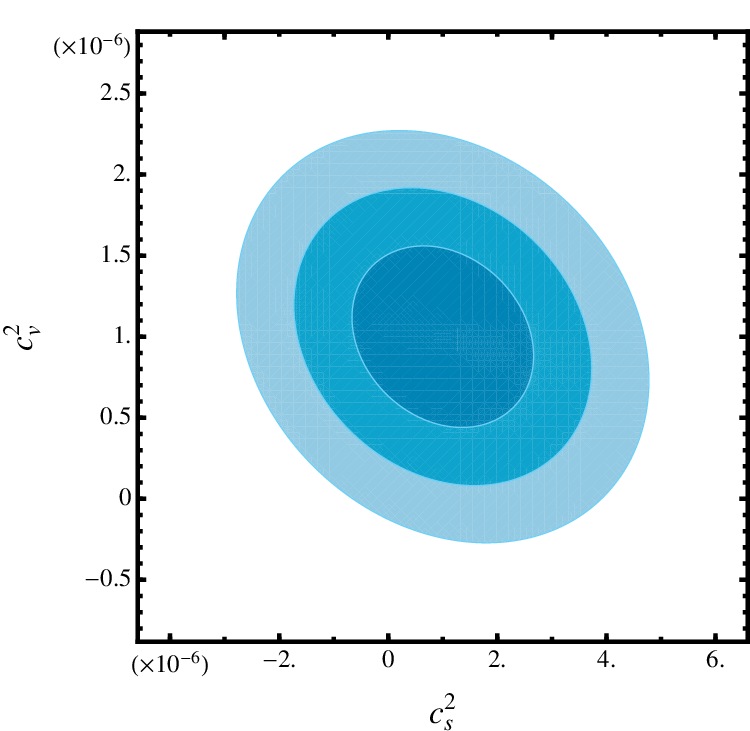,height=2.1in}
\epsfig{figure=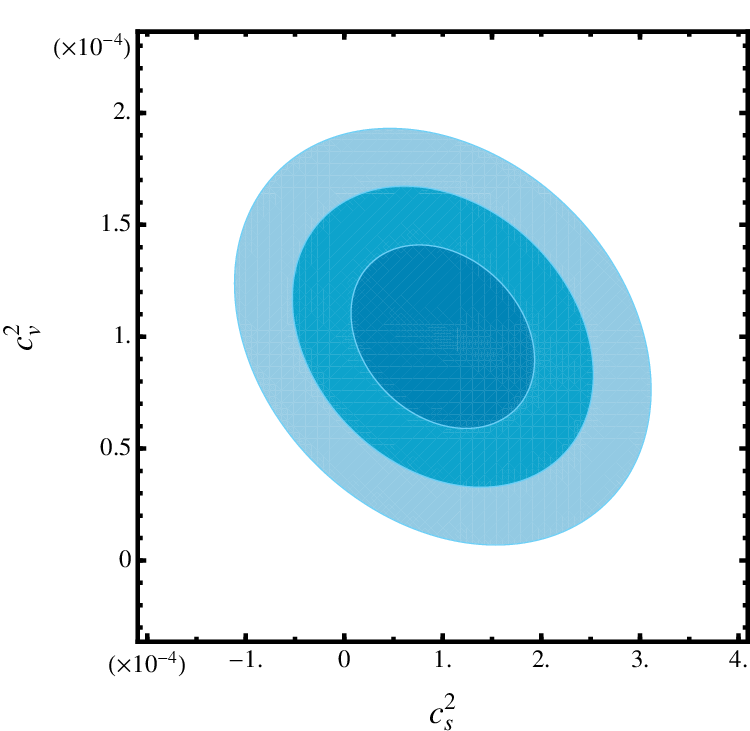,height=2.1in}
\epsfig{figure=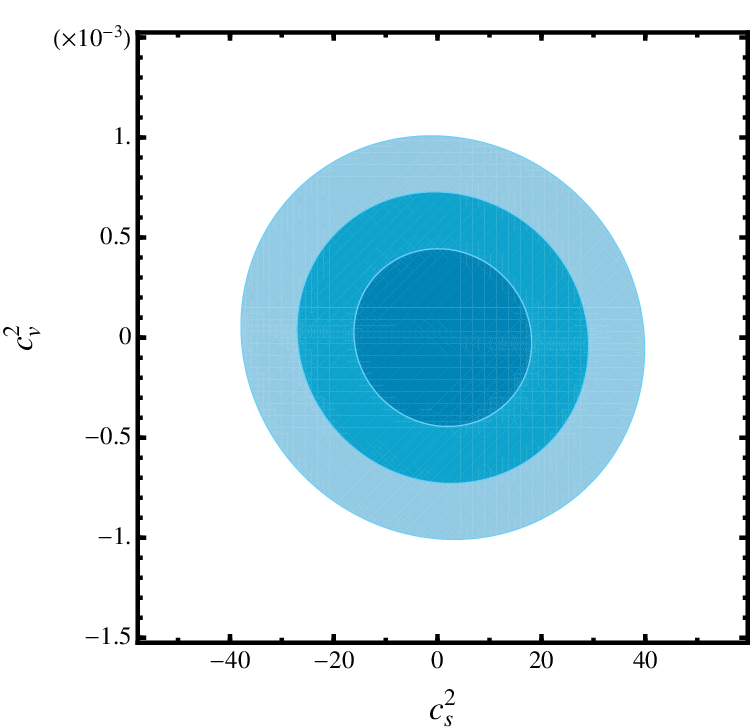,height=2.1in}
\caption{Errors on the sound speed and viscosity parameter from the Euclid galaxy redshift survey: 1, 2 and 3 $\sigma$ Fisher ellipses for the parameters $\cs$ and $\cv$ . The top panel corresponds to the fiducial $\cs = \cv = 10^{-6}$, the central panel to $\cs = \cv = 10^{-4}$ while the bottom one to the case $\cs = 1$ and $\cv = 10^{-6}$. Here we can see that no degeneracy is present, but errors are again very large, of the order of $\sim 100\%$ in almost all cases (exceptions are $\sigma_{\cv} \sim 10\%$ for $\cs = \cv = 10^{-6}$ and $\sigma_{\cv} = 1000\%$ for $\cs = 1$, $\cv = 0$).
}\label{fig:fish-pk-cs2cv2}
\end{figure}

The results are shown in Fig. \ref{fig:fish-pk-cs2cv2}, 
where the $1$, $2$ and $3\sigma$ forecasted
 contours are shown for all the analyzed cases.
In the cases $\cs = \cv = 10^{-6}$ and $\cs = \cv = 10^{-4}$ 
we notice that the errors on both parameters are best but 
still of the order of $100\%$ (only exception is $\sigma_{\cv} \sim 10\%$ for $\cs = \cv = 10^{-6}$). In the case where $\cs = 1$ and 
$\cv = 0$ the error on the sound speed is even larger, 
i.e. about $1000 \%$.

For a more quantitative insight we list the fully marginalized errors on $\cs$ and $\cv$ on Table \ref{tab:errorsGC}.

\begin{table}
\begin{centering}
\begin{tabular}{|c|c|c|c|}
\hline 
\multicolumn{4}{|c|}{GC}\tabularnewline
\hline 
\hspace{0.5cm}$\cs$\hspace{0.5cm}  & $\cv$  & $\sigma_{\cs}/\cs$ & $\sigma_{\cv}/\cv$ \tabularnewline
\hline 
$10^{-6}$  & $10^{-6}$  & $1.10 $  & $0.37 $  \tabularnewline
\hline 
$10^{-4}$  & $10^{-4}$  & $0.615$  & $0.271$ \tabularnewline
\hline 
$1$  & $0$  &  $11.3$ & $\sigma_{\cv} = 2.93 \times 10^{-4}$ \tabularnewline
\hline 
\end{tabular}
\par\end{centering}

\caption{Relative errors on the parameters $\cs$ and $\cv$ from the Euclid galaxy redshift survey. For the case $\cv = 0$ the absolute error $\sigma_{\cv}$ is given.}

\label{tab:errorsGC} 
\end{table}

\subsection{Combining weak lensing and galaxy clustering}

As we have seen previously,
 it seems that WL data, if one could remove their degeneracy, would provide quite good constraints to speed of sound and viscosity (or at least not much worse than GC does).
Hence adding WL data to GC observations should improve constraints on $\cs$ and $\cv$.

\begin{figure}
\centering
\epsfig{figure=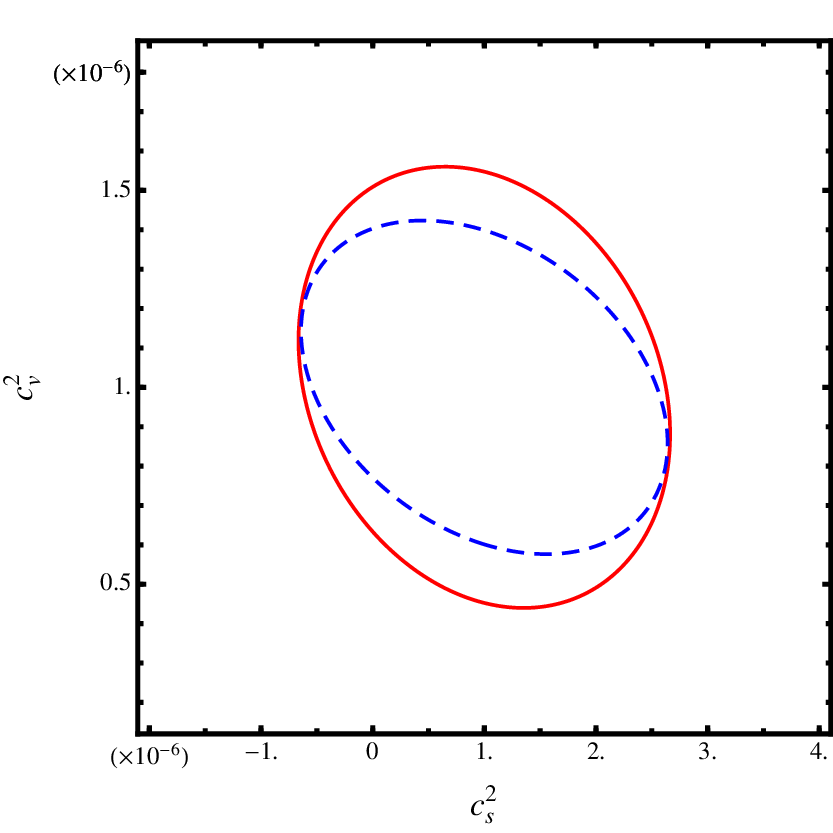,width=2.3in}
\epsfig{figure=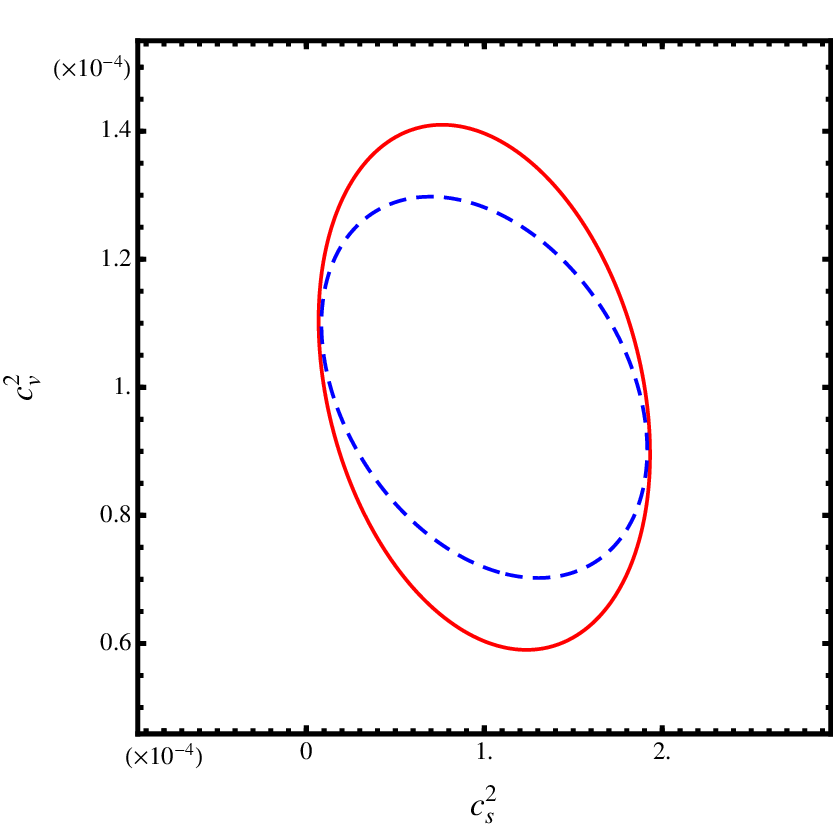,width=2.3in}
\epsfig{figure=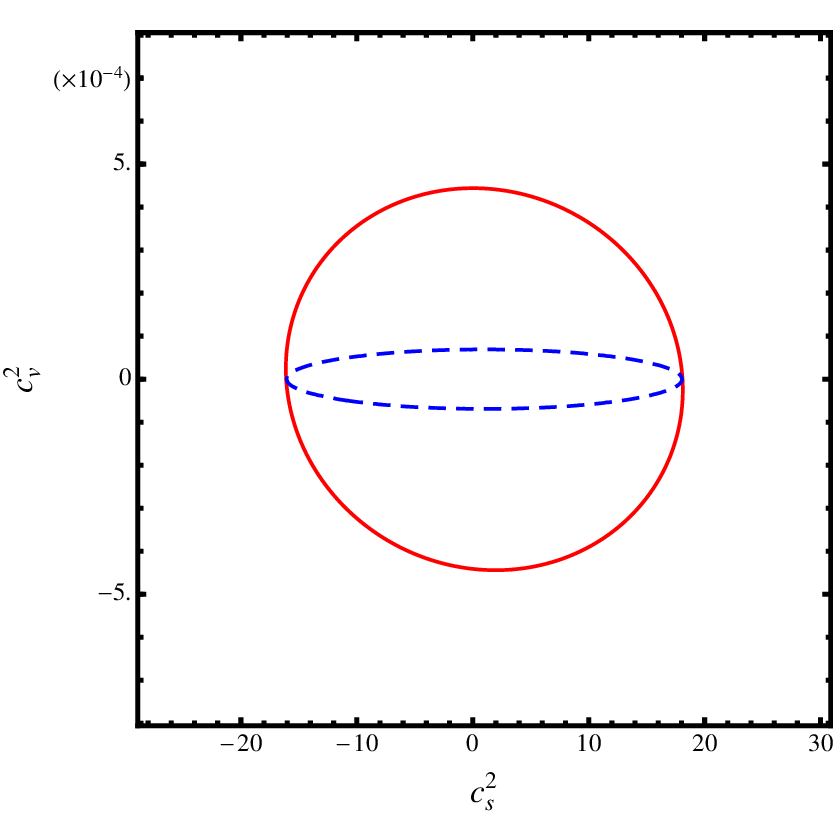,width=2.3in}
\caption{Errors on sound speed and viscosity parameter from GC (red solid ellipses) , and from the combination GC and WL data (blue dashed ellipses): $1\sigma$ Fisher ellipses for the parameters $\cs$ and $\cv$.  The top panel corresponds to the fiducial $\cs = \cv = 10^{-6}$, the central panel to $\cs = \cv = 10^{-4}$ while the bottom one to the case $\cs = 1$ and $\cv = 10^{-6}$. The Fisher matrices were produced by neglecting the covariance between WL and GC observations. It can be seen that the addition of WL data improves visibly the constraints, by mostly constraining $\cv$.
}\label{fig:fish-comb-zoom}
\end{figure}

The simplest way to produce forecasts on our combined data sets is
 to neglect the covariance between WL and galaxy 
 clustering observations. By doing so, we obtain the forecasted error 
 ellipses shown in Fig. \ref{fig:fish-comb-zoom}, where we also show
  the error ellipses from galaxy clustering only. Here we see that 
  WL observations reduce errors on the two parameters, 
  if not strongly at least visibly. We have already explained earlier 
  why WL observations give degenerate constraints on 
  $\cs$ and $\cv$ when these have very small fiducial values, 
  while  the main parameters observed by galaxy clustering are 
$Q$ and $G$ which are not as degenerate as $Q$ and $\Sigma$. 
Nevertheless, the reason why galaxy clustering observations 
give the main contribution in constraining $\cs$ and $\cv$, as can be 
seen by noticing the relatively small improvement in the 
combined errors in Fig.  \ref{fig:fish-comb-zoom}, probably lies 
only in the properties of the survey, and most likely the very small redshift error has a big role in this too.

When combining GC data to the more 
conservative WL data (computed using $k_{max} = 0.5 \,h/{\rm Mpc}$), the latter do not improve constraints. 
For this reason it will be very important in the future to build a
 reliable model for the non-linear regime, both for the standard
  $\Lambda$CDM model and for our imperfect fluid model.

We summarize the result of joining WL and GC data on the fully marginalized errors on $\cs$ and $\cv$ in Table \ref{tab:errorsWL+GC}.

A final remark to close this section: to produce a more accurate forecast of how the combination of WL and GC data will affect errors on $\cs$ and $\cv$ we would need to evaluate accurately the covariance between the two data sets. Although the number of galaxies observed spectroscopically is only $\sim 1/4$ of those observed photometrically, and we do not expect covariance to be very large, it could affect the estimate. This is left for future work.

\begin{table}
\begin{centering}
\begin{tabular}{|c|c|c|c|}
\hline 
\multicolumn{4}{|c|}{WL $+$ GC}\tabularnewline
\hline 
\hspace{0.5cm}$\cs$\hspace{0.5cm}  & $\cv$  & $\sigma_{\cs}/\cs$ & $\sigma_{\cv}/\cv$ \tabularnewline
\hline 
$10^{-6}$  & $10^{-6}$  & $1.08 $  & $0.28 $  \tabularnewline
\hline 
$10^{-4}$  & $10^{-4}$  & $0.604 $  & $0.197$  \tabularnewline
\hline 
$1$  & $0$  &  $11.3$ & $ \sigma_{\cv} = 4.5 \times 10^{-5}$  \tabularnewline
\hline 
\end{tabular}
\par\end{centering}

\caption{Relative errors on the parameters $\cs$ and $\cv$ from the  Euclid survey,  including  WL and  GC data and neglecting correlations between them. For the case $\cv = 0$ the absolute error $\sigma_{\cv}$ is given.}

\label{tab:errorsWL+GC} 
\end{table}

\section{Model comparison} \label{sec:model-comparison}

To understand more quantitatively whether Euclid will be able to distinguish viscous dark energy from other less exotic models, we estimate in this section the expected Bayesian evidence, which gives a measure of the probability of one model with respect to the other.

The Bayes' factor $B_{12}$ of models $\M_1$ and $\M_2$ is defined as the ratio of the model likelihoods through (see \cite{Kunz:2006mc})
\be
\frac{p(\M_1 |d)}{p(\M_2 |d)} = \frac{\pi(\M_1)}{\pi(\M_2)} B_{12},
\ee
where $p(\M_i |d)$ is the normalized posterior probability distribution of $\M_i$ and $\pi(\M_i)$ is the prior probability distribution for the model.
If we do not have any reason to prefer one model over the other before
we see the data, then $\pi(\M_1) = \pi(\M_2) = 1/2$ and
\be
B_{12} = \frac{p(\M_1 |d)}{p(\M_2 |d)}.
\ee
In the case of nested models, i.e. when $\M_1$ can be obtained from $\M_2$ by fixing the parameter(s) $\omega$ to $\omega*$, the expression of  $B_{12}$ can be simplified to give the Savage-Dickey density ratio (SDDR), see e.g. \cite{Trotta:2005ar}:
\be
B_{12}^{SDDR} = \frac{p(\omega | d, \M_2)}{\pi(\omega | \M_2)}\bigg|_{\omega = \omega^*},
\ee
where $p(\omega | d, \M_2)$ is the normalized posterior probability distribution and $\pi(\omega | \M_2)$ is the  prior probability distribution for the parameter $\omega$ of model $\M_2$, marginalized over all other parameters.

The ratio of the model probabilities can be interpreted as ``betting odds'', and since we set the prior probabilities of all models equal, the same is true
for the Bayes factors between models. It makes then sense to use logarithm of the Bayes factor, which we will call the Bayesian evidence,
\be
\mathcal{E} = \ln B_{12}.
\ee
We will use the interpretation of $\mathcal{E}$ in terms of Jeffrey's scale \cite{Jeff} as given in \cite{Trotta:2005ar}:
The evidence of one model against another one is denoted 
as not significant, substantial,
strong or decisive when $|\mathcal{E}|$ is $<1$, $1-2.5$, $2.5-5$ and $>5$, respectively.

In order to understand whether a positive viscosity can be distinguished from no viscosity by our data, let us take as our model $\M_1$ a k-essence fluid, i.e. a dark energy model where the sound speed is a free parameter. Hence we will have free $\cs$ while having fixed ${c_v^*}^2$ to zero. Our model $\M_2$ will be instead the full viscous dark energy model. For $\M_2$'s extra parameter $\cv$,
we choose a flat prior $\pi(\cv|\M_2)$  in the range $0\leq\cs\leq 1$, and $\M_1$ is nested in $\M_2$ at $c_v^2=0$. The motivation for choosing a positive value for $\cv$ is to avoid instabilities in the solution for the anisotropic stress function (see  \cite{fingerprinting3} for a more detailed explanation).
As our likelihood function we take the one forecasted with the Fisher matrix method: $p(d | x_i, \M_2) \propto \exp \left[ -1/2 (x_i-\bar x_i)^T F_{x_i x_j} (x_j-\bar x_j) \right]$, where $\bar x_i$ is the fiducial value of parameter $x_i$. We marginalize the likelihood function $p(d | x_i, \M_2) $ over all possible values of $\cs$ so that we obtain
\be
p(d | \cv, \M_2) = \frac{1}{N} \exp\left[- \frac{1}{2} \left(\frac{\cv-\bar c_v^2}{\sigma_{\cv}}\right)^2  \right],
\ee
where $\sigma_{\cv} $ is estimated with the Fisher matrix: $\sigma_{\cv} = (F^{-1})_{\cv\cv}$ 
and the normalization constant of the likelihood is $N = \sqrt{2\pi} \sigma_{\cv}$.
Our posterior $p(\cv | d, \M_2) \propto  p(d | \cv, \M_2) \pi(\cv|\M_2)$ is hence nonzero only in the interval $0\leq \cv\leq 1$, where it is given by
\be
p(\cv | d, \M_2) =
		\frac{1}{N'}  \exp\left[- \frac{1}{2} \left( \frac{\cv-\bar c_v^2}{\sigma_{\cv}}\right)^2  \right],
\ee
where the normalization constant $N'$ is computed by integrating $p(\cv | d, \M_2)$ over the interval $0\leq \cv\leq 1$.
The Bayesian evidence is therefore
\be
\mathcal{E} \equiv \ln B_{12}  = \ln \frac{\Delta \cv}{ N'} - \frac{1}{2}  \left(\frac{{c_v^*}^2-\bar c_v^2}{\sigma_{\cv}}\right)^2.
\ee
where $\Delta \cv$ is the interval in $\cv$ allowed by the flat prior, so that $\Delta \cv = 1$ in our case. 
Let us understand in detail the meaning of this equation. A positive $\mathcal{E}$ will indicate evidence in favor of $\M_1$, i.e.  k-essence, while a negative $\mathcal{E}$ will correspond to evidence in favor of $\M_2$, i.e. viscous dark energy. $|\mathcal{E}|$ indicates the strength of this evidence and can be compared to Jeffreys' scale, as explained above.
The term $-(1/2) ({c_v^*}^2-\bar c_v^2)^2/\sigma_{\cv}^2$ measures the distance between the model with posterior mean $\bar c_v^2$, in our case the chosen fiducial $\cv$, and the simpler model, in our case k-essence (${c_v^*}^2 = 0$). We see that this term is always negative, meaning that it contributes evidence in favor of model $\M_2$. It accounts for the fact that the ``true" (fiducial) model is fitting the data better (by construction in our case).
The term $\ln \Delta \cv /N'$ measures the ratio of the area of the $\cv$ parameter space allowed by the more complex model to the area of the error ellipse. It represents therefore the ``Occam razor'' term, disfavoring too complex models that ``waste'' too much of the prior space. Since in our case $\ln \Delta \cv = 0$, we are left with the term $-\ln N'$. This is the only term that can push the evidence towards positive values, in favor of model $\M_1$ i.e.  k-essence. When the simpler model is favored, it means that the improvement gained by having a better fit to the data is not compensating for the increased complexity of the model. 

Our results are summarized in Table \ref{tab:bayesianevidence-2}. We notice that the only case where there is (strong) evidence in favor of viscous dark energy corresponds to the choice of posterior mean $\bar c_s^2 = \bar c_v^2 = 10^{-4}$ and to the use of both WL and GC. In all the other cases the simpler k-essence is favored, so Euclid data alone are not enough to allow a detection of viscosity. An important reason for this result comes from our choice of prior which implies that we would have expected to measure any $\cv$ between 0 and 1 with equal probability. When the data favors a $\cv$ very close to zero then Occam's razor indicates that exactly zero is the better answer. Overcoming this effect then requires a highly significant detection of $\cv\neq0$. Of course in the case where the fiducial model has zero viscosity necessarily the simpler model is always favored, and having more data strengthens its evidence.

\begin{table}[tb]
\begin{centering}
\begin{tabular}{|c|c|c|c|}
\hline 
\multicolumn{4}{|c|}{Bayesian evidence $\mathcal{E} = \ln B_{12}$ }\tabularnewline 
\multicolumn{4}{|c|}{$\M_1$: fluid K-essence (variable $\cs$, $\cv = 0$)}\tabularnewline
\hline
\hspace{0.5cm}  \hspace{0.5cm}  &$\bar c_s^2 = 1$ $\bar c_v^2 = 0$  & $\bar c_s^2 = \bar c_v^2 = 10^{-4}$  & $\bar c_s^2 = \bar c_v^2 = 10^{-6}$ \tabularnewline
\hline 
GC  &  $7.9$  &  $ 2.8 $ &  $10.2$\tabularnewline
\hline 
GC $+$ WL  &  $9.8$ &  $-3.0$ & $7.8$  \tabularnewline
\hline 
\end{tabular}
\par\end{centering}
\caption{Bayesian evidence, $\ln B_{12}$, for the models $\M_1$ and $\M_2$, where $\M_1$ is a fluid k-essence model, with $\cs$ allowed to vary and no viscosity, while $\M_2$ is a viscous dark energy model. We compute $\ln B_{12}$ for different posterior means $\bar c_s^2$ and $\bar c_v^2$.} 
\label{tab:bayesianevidence-2} 
\end{table}

\begin{figure}
\centering
\epsfig{figure=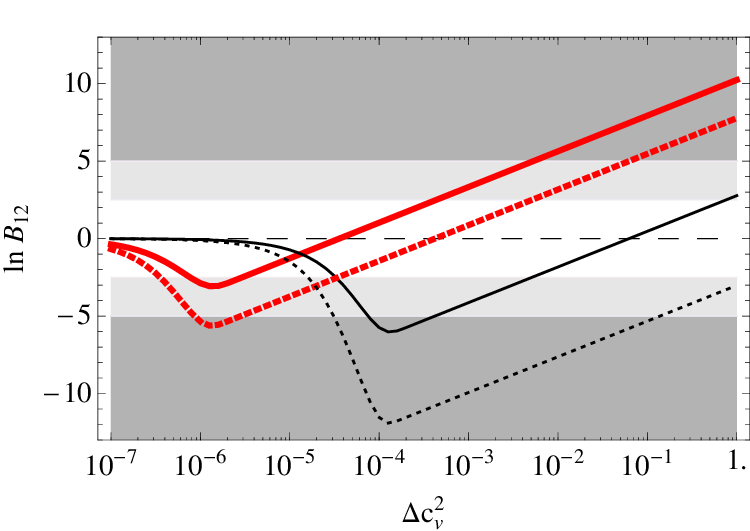,width=3.5in}
\caption{Bayesian evidence, $\ln B_{12}$, for the models $\M_1$  (a k-essence fluid with arbitrary constant sound speed $\cs$) and $\M_2$ (viscous dark energy), as a function of $\Delta \cv$, i.e. the range in $\cv$ allowed by our (flat) prior: $\Delta \cv = c^2_{v, max}-c^2_{v, min}$, where we fix $c^2_{v, min}$ to 0. The black thin (red thick) lines represent the case where the fiducial model (also best fit of the posterior) is $\bar c_v^2 = \bar c_s^2 = 10^{-4}$ ($\bar c_v^2 = \bar c_s^2 = 10^{-6}$). The solid (dashed) lines indicate the use of GC (GC and WL) data. The darker (lighter) shaded regions correspond to decisive (strong) evidence in favor of viscous DE (where $\ln B_{12} < 0$) or k-essence (where $\ln B_{12} > 0$).
We see that the evidence in favor of viscous DE increases until the $c^2_{v, max}$ of the prior reaches a value of the order of the fiducial $\cv$ considered, and then decreases as the log of $\Delta \cv$, reaching eventually evidence in favor of k-essence.
 In the case of using only GC data, we see that when choosing $\bar c_v^2 = \bar c_s^2 = 10^{-6}$, it is not possible to reach decisive but only strong evidence, while when $\bar c_v^2 = \bar c_s^2 = 10^{-4}$ decisive evidence is reached when $\Delta \cv \simeq 10^{-4}$. Adding WL data allows also decisive evidence in favor of the model with $\bar c_v^2 = \bar c_s^2 = 10^{-6}$ to be reached (if $\Delta \cv \simeq 10^{-6}$). Also, using joint GC and WL data, decisive evidence in favor of the viscous DE model with $\bar c_v^2 = \bar c_s^2 = 10^{-4}$ is reached for any prior having $\Delta \cv$ between $2\times 10^{-5}$ and $10^{-1}$.
}\label{fig:evidence-vs-prior}
\end{figure}

The effect of the prior is shown in Fig. \ref{fig:evidence-vs-prior}, where the black thin (red thick) lines indicate the case where the fiducial (true) model has $\bar c_v^2 = \bar c_s^2 = 10^{-6}$ ($\bar c_v^2 = \bar c_s^2 = 10^{-4}$). Solid (dashed) lines represent the use of GC (GC and WL) data.  Here we notice that, as we expect, when the range of the flat prior is of the order of the fiducial value of $\cv$ then we obtain the maximal evidence in favor of the viscous DE model. As soon as the interval allowed by the prior becomes much larger than the expected value for $\cv$, the more complex model is disfavored as it wastes too much prior space, and the evidence in favor of the simpler k-essence model starts growing (as the log of $\Delta \cv$) until the k-essence model becomes favored.  In particular we see that, when using both GC and WL, the evidence in favor of viscous DE becomes decisive in the case $\bar c_v^2 = \bar c_s^2 = 10^{-4}$ if the prior is $2\times 10^{-5}\lesssim \Delta \cv\lesssim 10^{-1}$, in the case $\bar c_v^2 = \bar c_s^2 = 10^{-6}$ only in the narrower range $8\times 10^{-7}\lesssim\Delta \cv\lesssim  2 \times10^{-6}$. We finally note that if the prior range is much smaller than the fiducial value, then the posterior (which is limited by the prior) becomes flatter and flatter so that the Bayes' factor converges towards unity for very small priors.

We could also have compared standard quintessence ($\cs =1$ and $\cv = 0$) to our viscous dark energy model. However, for all cases with small sound speeds we would  have obtained a very decisive evidence in favor of viscous dark energy -- but the result would have been driven by the detection of a sound speed different from the speed of light, rather than by a non-zero viscosity: as shown in \cite{fingerprinting2}, it is possible to clearly distinguish the true sound speed from $c_s=1$ for $c_s \lesssim 10^{-2}$ even in the absence of viscosity (at least for our choice of $w=-0.8$). The use of k-essence as the comparison model allows to focus exclusively on the question whether a non-zero viscosity could be detected.

\section{Conclusions} \label{sec:conclusions}

In this work we have studied how well a viscous dark energy
 model can be constrained by the Euclid weak lensing and galaxy
 clustering surveys.

The model was first proposed by \cite{Hu:1998kj} to describe a fluid with anisotropic stress due to viscosity. It can reproduce the neutrino anisotropy up to the quadrupole, and setting the anisotropy parameter $\cv$ to zero reduces it to the standard case of no viscosity. Our dark energy imperfect fluid is hence parameterized by $\cv$, together with the speed of sound and the background equation of state.

The Euclid survey is in principle particularly apt to constrain our model because -- through both its WL and galaxy 3D power spectrum probes -- it measures, together with the background expansion, also the metric perturbations, which is essential to constrain perturbation quantities such as the viscosity and the speed of sound.

As regards WL, before forecasting the errors with the Fisher matrix technique, we have analyzed the WL power spectrum. We have found that it constrains the clustering parameter $Q$, the WL potential $\Sigma$ and the growth factor $G$. Of these, the most sensitive to $\cs$ and $\cv$ are $\Sigma$ and $Q$. For very small values of the speed of sound and the viscosity, their dependence on these parameters is identical at most scales. This generates a problem for WL constraints, as they will be almost completely degenerate. This is evident once we compute the forecasted errors. We do this for three fiducial models: $\cs = \cv = 10^{-6}$, $\cs = \cv = 10^{-4}$ and $\cs = 1$ $\cv = 0$ but due to the degeneracy problem the only reliable errors are those computed for the last case and amount to $\sigma_{\cs} = 201$, $\sigma_{\cv} = 1.55 \times 10^{-2}$. The previous results correspond to the optimistic case of having a reliable model for the non-linear modes up to $\ell = 5000$. If we use the more realistic limiting $\ell = 30$, the degeneracy is weakened -- because less of the degenerate scales are used -- and we can also trust results for the case $\cs = \cv = 10^{-4}$, but the errors get even worse.

As regards the galaxy power spectrum, it constrains $Q$, $G$ and the growth rate $f$ through redshift space distortions. The most sensitive parameters to the speed of sound and the viscosity are here $Q$ and $G$, and, contrary to the WL case, there is no degeneracy problem. Using the Fisher matrix method, we forecast errors on our model's parameters. They are of the order of $100\%$ for both $\cs$ and $\cv$ for the fiducial models $\cs = \cv = 10^{-6}$ and $\cs = \cv = 10^{-4}$, while they are of about $1000\%$ on $\cs$ for the fiducial case $\cs = 1$, $\cv = 0$.

If we combine the optimistic WL forecast and GC constraints (assuming no covariance between these data) we improve visibly our results, while when adding the  more pessimistic WL case to galaxy power spectrum constraints are left almost unchanged.

To quantify the ability of Euclid to distinguish viscous dark energy from models with no viscosity, we evaluate the Bayesian evidence and find that for the fiducial model with $\cs= \cv = 10^{-4}$ the joint GC and WL survey will be able to provide strong evidence in favor of viscosity, while in all other cases the higher complexity of the model is not compensated by a better fit to the data. This is partially due to our choice of uniform prior on $\cv$ in the range $[0,1]$. If we relax this assumption we find that it is possible to obtain decisive evidence in favor of viscous DE even with GC data alone if the range of $\cv$ allowed by the prior is of the order of the fiducial $\cv$, and for a wider choice of prior range when using both GC and WL data and for the case $\bar c_v^2 = \bar c_s^2 = 10^{-4}$.

Summarizing, future galaxy surveys like Euclid will only marginally be able to constrain viscous dark energy models.
This statement depends of course on the model being analyzed. Some modified gravity models have a larger
effective anisotropic stress than the viscous dark energy models considered here, and Euclid may have a better chance
at detecting a non-zero $\sigma$ in this case.
Alternatively one could combine Euclid with other observations such as the cosmic microwave background, cluster counts, data from type-Ia supernovae and ISW measurements: all these observable constrain themselves $\cs$ and $\cv$ and/or reduce the errors on background parameters and therefore reduce also marginalized errors on perturbation quantities.
These interesting developments are left to future work.


\begin{acknowledgments}
We thank Licia Verde and Juan Garcia-Bellido for useful discussions. 
DS acknowledges support from the JAEDoc program with ref. JAEDoc074 
and the Spanish MICINN under the project AYA2009-13936-C06-06.
DS also acknowledges financial support from the Madrid Regional Government 
(CAM) under the program HEPHACOS P-ESP-00346, Consolider-Ingenio 2010 PAU (CSD2007-00060), 
as well as in the European Union Marie Curie Network ``UniverseNet" under contract MRTN-CT-2006-035863.
EM was supported by the Spanish MICINNs Juan de la Cierva programme (JCI-2010-08112), 
by CICYT through the project FPA-2009 09017, by the Community of Madrid  through the project 
HEPHACOS (S2009/ESP-1473) under grant P-ESP-00346 and by the European Union FP7  ITN INVISIBLES (Marie Curie Actions, PITN- GA-2011- 289442). 
MK acknowledges funding by the Swiss NSF.

\end{acknowledgments}

\end{document}